\documentclass[journal,12pt,onecolumn,draftclsnofoot]{IEEEtran}

\usepackage[colorlinks,bookmarksopen,bookmarksnumbered,citecolor=red,urlcolor=red]{hyperref}
\usepackage{mathtools}
\usepackage{cuted}
\usepackage{epsfig}
\usepackage{float}
\usepackage{amsmath}
\usepackage{amssymb}
\usepackage{amsmath}
\usepackage{mathrsfs}
\usepackage{color}
\usepackage{subfigure}
\usepackage{graphicx}
\usepackage{anyfontsize}
\usepackage{t1enc}
\usepackage{cite}
\usepackage{epstopdf}
\usepackage{tabularx,booktabs,caption}
\usepackage{algorithm,algpseudocode,authblk}
\usepackage{textcomp}

\ifCLASSINFOpdf
\else
\fi

\begin{document}
	\title{{Performance of UAV assisted Multiuser Terrestrial-Satellite Communication System over Mixed FSO/RF Channels}}
	\author{
		\IEEEauthorblockN{Praveen Kumar Singya and Mohamed-Slim Alouini, \IEEEmembership{Fellow, IEEE}}
		\thanks{P. K. Singya and M.-S. Alouini are with the Computer, Electrical, and Mathematical Science and Engineering (CEMSE) Division, King Abdullah University	of Science and Technology (KAUST), Thuwal 23955-6900, Saudi Arabia (e-mail:praveen.singya@kaust.edu.sa, slim.alouini@kaust.edu.sa)}
		\vspace{-2em}
	}
	\maketitle
	
	\begin{abstract}
		In this work, performance of a multi-antenna multiuser unmanned aerial vehicle (UAV) assisted terrestrial-satellite communication system over mixed free space optics (FSO)/ radio frequency (RF) channels is analyzed. Downlink transmission from the satellite to the UAV is completed through FSO link which follows Gamma-Gamma distribution with pointing error impairments. Both the heterodyne detection and intensity modulation direct detection  techniques are considered at the FSO receiver. To avail the antenna diversity, multiple transmit antennas are considered at the UAV. Selective decode-and-forward scheme is assumed at the UAV and opportunistic user scheduling is performed while considering the practical constraints of outdated channel state information (CSI) during the user selection and transmission phase. The RF links are assumed to follow Nakagami-m distribution due to its versatile nature.
		In this context, for the performance analysis, analytical expressions of outage probability, asymptotic outage probability, ergodic capacity, effective capacity, and generalized average symbol-error-rate expressions of various quadrature amplitude modulation (QAM) schemes such as hexagonal-QAM, cross-QAM, and rectangular QAM are derived. A comparison of various modulation schemes is presented. Further, the impact of pointing error,  number of antennas, delay constraint, fading severity, and imperfect CSI are highlighted on the system performance. Finally, all the analytical results are verified through the Monte-Carlo simulations.
	\end{abstract}
	\begin{IEEEkeywords}
	Satellite, multiantenna UAV,  mixed FSO/RF, Gamma-Gamma, pointing errors, outdated CSI, average symbol-error-rate (ASER), hexagonal QAM (HQAM), cross QAM (XQAM), rectangular QAM (RQAM).
	\end{IEEEkeywords}
	\section{Introduction}	
	Recently, readily deployable flying wireless access devices like unmanned aerial vehicles (UAVs)  have gain an increased attention for robust and reliable communication in various applications including military operations, temporary social events or disastrous situations to provide improved coverage, capacity, and data-rates to meet the requirements of $5^{th}$ generation (5G) and beyond communications \cite{mozaffari2019tutorial,zhan2011wireless}. 
	After incredible efforts from the industries and academia, it has been found that the terrestrial networks alone are not able to provide the desired data-rates targeted for the 5G and beyond communication systems and satellite communication has emerged as an essential approach \cite{guidotti2019architectures}. However, in practice, masking occurs between the satellite and the terrestrial users and line-of-sight (LoS) communication becomes  difficult. This can be solved by deploying the terrestrial ground relays which can transmit the satellite signal to improve the throughput and reliability of the satellite communication system. However, such infrastructures may damaged in some disastrous situations such as in flood conditions or in earthquake situations. Also, in case of some temporary events, readily deployable infrastructures such as UAVs are very helpful to provide reliable wireless communication over the hybrid satellite-terrestrial networks (HSTNs) \cite{sharma2019outage}.
	{On the basis of operating range and size, platforms for UAV are categorized into low altitude platform (LAP), suitable for only few hundred meters and high altitude platform (HAP), ranges in kilometers \cite{sharma2019random}. Various applications and technological trends of HAP can be seen in \cite{d2016high}. Application of LAP in aerial-terrestrial communication can be seen in \cite{kandeepan2014aerial}.}
	Recently, UAVs have gained an increased attention 
	as flying base stations.
	In \cite{zeng2016throughput,ono2016wireless,choi2014energy}, the authors have considered the UAVs as the mobile relay nodes for the communication.
	In \cite{sharma2019random}, the authors have proposed the mobility modeling of random 3-dimensional (3D) mobile UAV networks and  investigated their coverage probability.
	However, the above mentioned works have not been considered in HSTNs.
	The UAV assisted HSTNs have received an increased attention and European ABSOLUTE project has adopted this in emergency situations for communication \cite{grace2013integrated}.
	In \cite{sharma2019outage}, 3D mobile UAVs assisted HSTN is considered and outage probability is derived for the performance analysis.
	In \cite{liu2019performance}, a UAV assisted multiuser HSTN is considered and outage and asymptotic outage probability are derived.
	However, \cite{sharma2019outage, liu2019performance} consider only the radio-frequency (RF) links for communication. 

	Recently, the high data-rate requirement with increased capacity and bandwidth  direct us towards the free space optics (FSO) as an efficient 
	replacement to the RF links. This is because of its easy deployment, low cost, and point-to-point high data-rate communication which provides high bandwidth  and operates in license free band \cite{singya2020performance,trichili2020roadmap}. 
	However, some challenges such as atmospheric turbulence, modulation, and pointing error need to be considered during the satellite to UAV optical communication  \cite{kaushal2016optical}. 
	For the modeling of FSO link with atmospheric turbulence, Gamma-Gamma distribution is preferred commonly due to its suitability to consider both the large and small scale atmospheric fluctuations.
	
	There are only few works which consider an optical link between the satellite and UAV.
	{In \cite{antonini2006feasibility}, feasibility of a satellite to HAP FSO link for the relay purpose is analyzed by obtaining the capacity and efficiency.}
	In \cite{li2018investigation}, authors have proposed an FSO based UAV to satellite optical communication system and the PDF expression and bit-error-rate (BER) expression for binary phase shift keying (BPSK) are derived for both the uplink and downlink scenario.
	In \cite{kong2020multiuser}, a multiuser UAV assisted decode-and-forward (DF) based HSTN is considered, where Gamma-Gamma fading for FSO link and correlated Rayleigh fading for RF links are considered and only the ergodic capacity is obtained.
	In \cite{liu2020outage}, a single user UAV assisted  amplify-and-forward (AF) based HSTN is considered, where Gamma-Gamma fading for FSO link with pointing error and correlated Rayleigh fading for RF link with imperfect channel state information (CSI) are considered and only outage probability is obtained.
	{In \cite{swaminathan2021haps},  a HAP  based uplink FSO/RF ground to satellite communication system is presented,  where Gamma-Gamma fading for FSO link with pointing error and shadowed-Rician fading for RF link is considered and the outage probability, asymptotic outage probability, and BER for BPSK is obtained.}

However, a considerable works on mixed RF/FSO or FSO/RF relay systems have been reported in the literature.
	In \cite{bhatnagar2013performance}, an AF based hybrid RF/FSO satellite-terrestrial communication system is considered, where Shadowed-Rician and Gamma-Gamma distributions are considered for the RF and FSO links, respectively. For the performance analysis, analytical average symbol-error-rate (ASER) expression for M-ary PSK (MPSK) is derived and asymptotic analysis is performed on it.
	In \cite{anees2015performance1} and \cite{anees2015performance}, DF and AF based asymmetric RF/FSO communication systems are considered, respectively, where FSO link is considered to have atmospheric turbulence influenced Gamma-Gamma fading with pointing error impairments and  RF link is Nakagami-m distributed. For the performance analysis,  analytical expressions of outage probability, average channel capacity, and average BER (ABER) for various MPSK and M-ary quadrature amplitude modulation (MQAM) are derived.
	In \cite{anees2015information}, AF based mixed RF/FSO system is considered,  where FSO link is considered to have atmospheric turbulence influenced Gamma-Gamma fading with pointing error impairments and  RF link is Nakagami-m distributed. For the performance analysis,  average channel capacity for optimum  simultaneous power and rate adaptation and truncated channel inversion with fixed rate schemes are derived.     
	In \cite{zedini2016performance}, an AF based hybrid FSO/RF system is considered where Gamma-Gamma and Nakagami-m distributions are considered for the FSO and RF links, respectively. For the performance analysis, outage probability, ergodic capacity, and BER of BPSK are obtained. In \cite{chen2019novel}, for a mixed FSO/RF system, energy harvesting is performed and outage probability expression is derived.
	In \cite{ahmad2017performance}, authors have consider a mixed FSO/RF satellite communication system where an uplink optical feeder link is considered to satellite and downlink RF links are considered to terrestrial users. For the performance analysis, outage probability, ergodic capacity, and ABER expressions are derived. 
	In \cite {zedini2020performance}, authors have considered a mixed FSO/RF satellite communication system where an uplink optical feeder link is considered to satellite which is Gamma-Gamma distributed  and downlink RF links are considered to terrestrial users  following the shadowed-Rician fading. Further, outage probability, ergodic capacity, and ABER expressions are derived for the performance analysis. {In \cite{yang2018performance}, a LAP based multihop RF/FSO/RF system is proposed and only the outage probability is obtained for the considered system.}  In \cite{altubaishi2019capacity}, capacity of an AF based multihop FSO/RF system is calculated.
	{In \cite {li2019performance}, for a nonorthogonal multiple access based multiuser hybrid FSO/RF system, outage probability and ergodic sum-rate are obtained. Performance of  a switching-based hybrid FSO/RF DF relay system with maximal-ratio combining (MRC) at the destination is analyzed in \cite{sharma2019switching}.
		In \cite{lee2020throughput}, for a UAV assisted hybrid FSO/RF system with finite size buffer, data throughput is maximized.}

	On the other hand, the family of QAMs has attracted significant attraction in present and future wireless communication systems due to its capability to provide bandwidth efficient high data-rate communication. The family mainly includes the square QAM (SQAM), rectangular QAM (RQAM), cross QAM (XQAM), and hexagonal QAM (HQAM) constellations. SQAM is valid only for the even power of 2 constellations. However, odd power of 2 constellations such as RQAM are also required for better channel adaptation. From average power point of view, RQAM is not a good choice and instead XQAM is preferred which has considerably lower peak and average powers than RQAM \cite{zhang2010exact}. However, the demand of high data-rates leads the research for more compact two dimensional (2D) constellations and a hexagonal lattice based constellation, HQAM comes into existence. HQAM has the densest 2D packing for a given Euclidean distance between the constellation points, hence, has the lower peak and average powers and provides optimum SER performance than the other QAM constellations \cite{singya2017impact,kumar2017aser,singya2019performance}.  HQAM is categorized into regular and irregular HQAM constellations, however, in this work, we have considered only the irregular HQAM constellations due to its optimum performance \cite{abdelaziz2018triangular,singya2020performance}.
	A significant work on the ASER performance of various QAM constellations including the HQAM have been reported in the literature \cite{singya2017impact,kumar2017aser,singya2019performance,shaik2019impact,shaik2019performance,parvez2019aser,abdelaziz2018triangular}. However, such works focus on the RF systems only. For the very first time, authors in \cite{singya2020performance} have investigated the ASER performance of various QAM constellations in a mixed RF/FSO system. 

	Motivated with this, in this work, a UAV {(HAP)} assisted  multiuser multiantenna  HSTN is considered, where the downlink transmission from the satellite to UAV is completed through the FSO link. A selective DF scheme is adopted at the UAV and the decoded signal is then broadcasted to the terrestrial users via RF links. It is assumed that the FSO link is modeled with the atmospheric turbulence influenced Gamma-Gamma fading with pointing error impairments and both the intensity modulation direct detection (IM/DD) and heterodyne detection techniques are considered at the FSO receiver. The RF links are assumed to follow the Nakagami-m distribution. Further, opportunistic user scheduling is performed for the multiuser diversity and outdated CSI is considered during the user selection and signal transmission due to the time varying nature of the channel statistics. Practical path-loss modeling is also considered for both the FSO and RF links. From this prospective, the major contributions of this work are as follow:
	\begin{itemize}
		\item For the considered system model, analytical outage probability expression is derived. The impact of FSO receiver's type, pointing error, correlation coefficient, fading severity of the RF link, and number of transmitting antennas at the UAV are shown on the performance of the outage probability.
		
		\item {To obtain the diversity order of the considered system, asymptotic outage probability is also obtained by performing the high signal-to-noise-ratio (SNR) approximation on the outage probability and the combined impact of number of users, number of transmit antennas, outdated CSI, fading coefficient, pointing error, atmospheric turbulence, and the type of FSO detector are observed on the system performance.}
		
		\item  Based on the instantaneous received rate at the UEs, ergodic capacity expression is derived. The impact of FSO receiver's type, pointing error, correlation coefficient, fading severity of the RF link, and number of transmitting antennas at the UAV are shown on the performance of the ergodic capacity.
		\item To guarantee a minimum delay and to maintain an efficient constant data-rate, effective capacity is also derived and the impact of delay constraint ($\Theta$) is observed on the performance of effective capacity.
		
		\item Based on the derived outage probability expression, a CDF based ASER analysis is performed for various generalized higher order QAM constellations including the general order HQAM, RQAM, and XQAM constellations.
		\item  For such complex higher order QAM constellation heterodyne detection at the FSO receiver is considered and the impact of pointing error, correlation coefficient, fading severity of RF link, number of transmitting antennas at the UAV, and number of terrestrial users are shown on the ASER performance of the general order HQAM, SQAM, RQAM, and XQAM constellations. 
		\item The ASER performances of various QAM constellations are compared which validate the superiority of the  HQAM constellation over the others. 
		\item Finally, all the analytical results are verified through the simulation results.
	\end{itemize}
	
	The rest of the paper is organized  as: Section II discusses the system and channel model in details. {Based on the end-to-end (e2e) instantaneous received SNR, outage and asymptotic outage probability are derived in Section III.} In Section IV, ergodic capacity and effective capacity are analyzed. Section V discusses the theoretical and simulation results in details. Finally, conclusions from the obtained results are drawn in Section VI.
	%
	\section{System and Channel Model}
	\begin{figure}[h!]
		\centering
		\includegraphics[width=2.5in,height=3.2in]{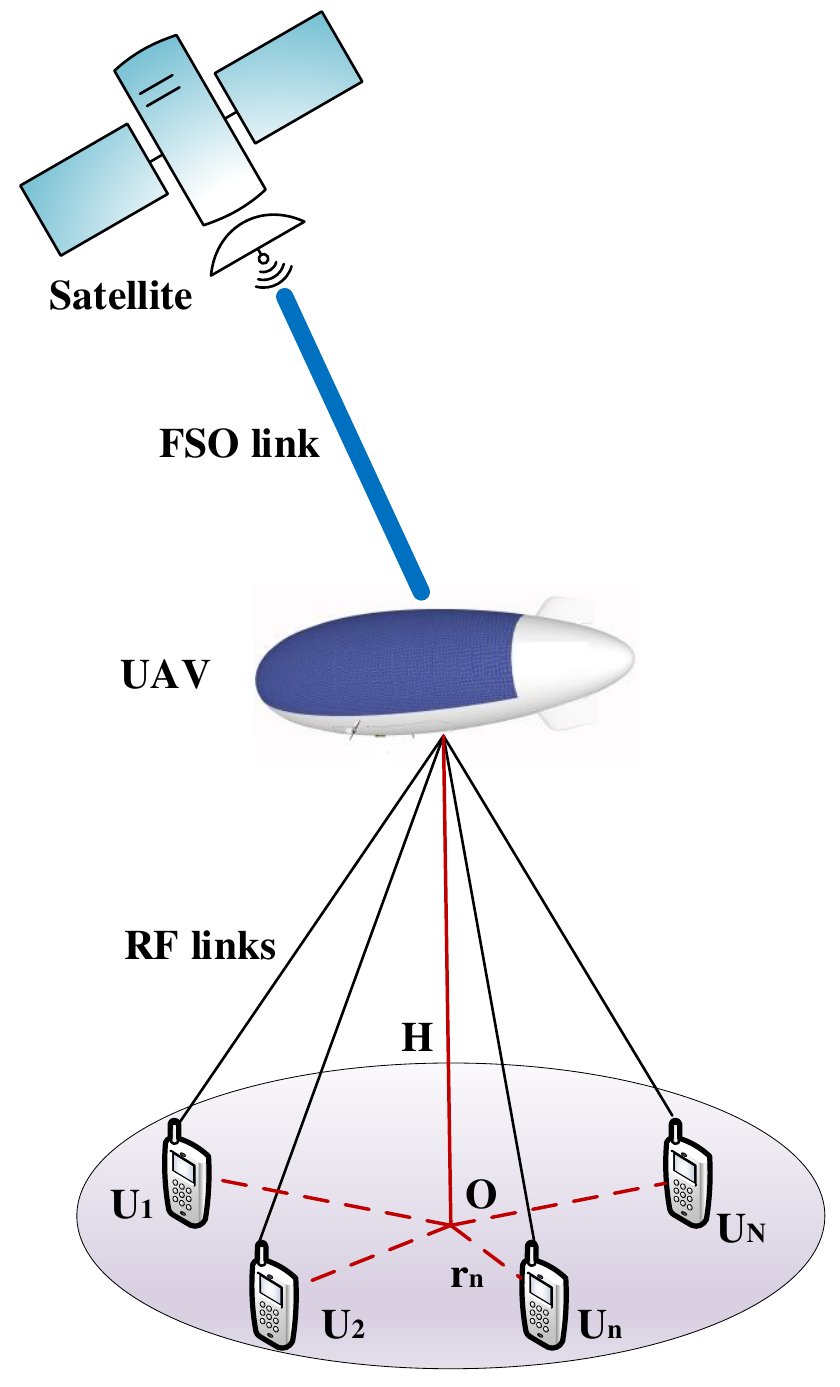}
		\caption{Considered system model.}
		\label{syst}
	\end{figure}
	In this work, a downlink asymmetric system model  is considered where multiples terrestrial user equipments (UEs) received information signal through the satellite (S) via a UAV  as shown in Fig. \ref{syst}. The N terrestrial UEs ($U_1, U_2,...,U_N$) with single antenna are considered and multiple antennas are deployed at the UAV. A significant path loss between the S and UAV exist and hence, a selective DF approach (threshold based DF relaying) is considered. An asymmetric FSO/RF communication model is considered for  e2e communication, where an FSO link between the satellite and UAV is considered and RF links between the UAV and multiple UEs are considered. The FSO link is assumed to be atmospheric turbulence induced Gamma-Gamma distributed with pointing error impairments. RF links are assumed to be Nakagami-m distributed.
	The e2e communication completes in two time phases.
	In the first time phase, satellite source (S) transmits an optical signal $x(t)$  ($\mathbb{E}[|x(t)|^2=1]$) to the UAV. Here, $\mathbb{E}[\cdot]$ represents the statistical expectation operator. {We have considered a 1550nm window for optical beam which is the most suitable one because it is eye safe and the Doppler effect is minimal than the other frequencies. At the LEO satellite, semiconductor laser is embarked and a lens collimates its beam to a 0.15m diameter telescope. At the receiver, a telescope with a lens diameter 0.25m is considered to receive the optical beam which is then passed through a filter for optical background noise reduction. We consider that a 10 Gbps signal is transmitted in optical domain with 20 GHz bandwidth \footnote{In \cite{trichili2020roadmap}, authors have shown a diverse range of data-rates (in tens of Gbps) of various commercial FSO systems for a particular distance (in km). Further, \cite{antonini2006feasibility} proposes a low earth orbit (LEO) satellite to HAP optical link setup capable of providing Gbps data-rates.}. Let suppose the Doppler shift may increase the bandwidth by 10 GHz, then we required nearly 30 GHz bandwidth for the filter. Hence, the optical bandwidth $B_0=30$ GHz is considered.} At the UAV, photodetector converts the  optical signal into electrical signal which results in 
	\begin{align}
		y_R(t)=({P_s}\zeta_R\eta I)^{i/2}x(t)+v_R(t),
	\end{align}
	where $I=I_pI_a$ represents  the FSO channel fading coefficient wherein $I_p$ is the pointing error and $I_a$ is the atmospheric turbulence induced fading. $\eta$ represents the optical-to-electrical conversion coefficient and $v_R(t)$ represents the additive white Gaussian noise (AWGN) with 0 mean and $\sigma^2_R$ variance, associated with the S-UAV link. The noise variance is $\sigma^2_R=\kappa B_o T$, where $\kappa=1.38\times10^{-23}$ represents the Boltzmann constant, $B_o$ is the optical receiver bandwidth, and $T$ is the noise temperature in kelvin. Also, $i$ represents the type of optical detector at the FSO receiver, where $i=1$ stands for the heterodyne detection and $i=2$ stands for the IM/DD detection. Further, $\zeta_R=1/2(G_S+G_R-A_{atm}-A_{FS}-L_{lenses}-M_s)$ dB is the amplitude of the path-loss  between the satellite and UAV, where $G_S$ and $G_R$ are the transmitter and receiver gains at the satellite and UAV, respectively,  $A_{FS}$ and $A_{atm}$ are the free-space and atmospheric attenuations, respectively,  $L_{lenses}$ is the lenses loss, and $M_s$ is the system margin \cite{antonini2006feasibility,kong2020multiuser}.{ The optical transmitter and receiver gains at the satellite and the UAV are  $G_S=\frac{\pi^2D_S^2}{\lambda_f}$ and $G_R=\frac{\pi^2D_R^2}{\lambda_f}$, respectively, where $D_S$ and $D_R$ are the lens diameters and $\lambda_f$ is the wavelength of the FSO link.}
	The instantaneous received SNR at the UAV is given as
	\begin{align}
		\gamma_R=\frac{({P_s}\zeta_R\eta I)^i}{\sigma^2_R}.
	\end{align}
	
	Selective DF relaying is preferred at the UAV. Hence, information signal at the UAV is first decoded successfully and is forwarded to terrestrial UEs only if  $\gamma_R$ is greater than or equal to a predefined threshold SNR ($\delta_{th}$). {If $\gamma_R< \delta_{th}$, UAV cannot decode anything from the received signal and no information is forwarded to the UEs. To avoid this, transmit signal power is maintained in such a way that the received SNR will always be greater than or equal to   $\delta_{th}$, so that the system will not be in outage.} 
	Hence, $\phi(\gamma_R)$ is defined as
	\begin{align}
		\phi(\gamma_R)=
		\begin{cases}
			0,  & \text{for }  \gamma_R < \delta_{th}\\
			1,  & \text{for }  \gamma_R\geq \delta_{th}
		\end{cases}
	\end{align}
	
	During the second communication phase, decoded electrical signal at UAV is  transmitted to the terrestrial UEs.  To improve the transmission reliability and to improve the channel capacity, $N$ transmit antennas are considered at the UAV. Further, single antenna is considered at the UEs due to size limit constraint.
	Thus, after transmit beamforming, received signal at the $n^{th}$ UE is
	\begin{align}
		y_{n}(t)=\phi(\gamma_R)\sqrt{P_R} \zeta_n \textbf{h}^{H}_{n} \textbf{w}_n x(t) + v_2(t),
	\end{align}
	where $P_R$ is the transmit power at the UAV.   $\textbf{w}_n \in \mathbb{C}^{N_t\times 1}$ is the transmit beamforming weight vector and $\textbf{h}^{H}_{n}$ is the $N_t\times 1$ channel vector between the UAV and the $n^{th}$ UE. Here, $(\cdot)^H$ represents the Hermitian operator. The beamforming weight vector $\textbf{w}_n=\frac{\textbf{h}_n}{||\textbf{h}_n||_F}$ is chosen according to the maximal ratio transmission (MRT) principle, where $||(\cdot)||_F$ represents the Frobenius norm. Also, $v_2(t)$ denotes the AWGN with 0 mean and $\sigma^2_n$ variance, associated with the UAV-$U_n$ link. Here, $\sigma^2_n=\kappa B_r T$, where $B_r$ is the noise bandwidth at the RF UEs. Further, $\zeta^2_n=\frac{1}{2}\big(20~ \text{log}(\lambda_{RF})-20\alpha_t~ \text{log}(d_k)-20~\text{log}(4\pi)\big)$ dB represents the path loss between the UAV and the $n^{th}$ UE, where $d_k$ is the distance of the $n^{th}$ UE from the UAV, $\lambda_{RF}$ is the wavelength of the RF link, and $\alpha_t\in[2,4]$ is the path loss factor of the RF link \cite{huang2019performance}.

	Thus, the e2e instantaneous SNR at the $n^{th}$ UE is given as 
	\begin{align}
		\gamma_{U_n}=\phi(\gamma_R)\frac{P_R \zeta_n^2|\textbf{w}^H_n\textbf{h}_{n}|^2}{\sigma^2_n}
		=\phi(\gamma_R)\gamma_{RU_n}.
	\end{align}
	
	The satellite to UAV link is an FSO link, its fading channel coefficient is given as $I=I_pI_a$.
	The atmospheric turbulence $I_a$ is modeled with Gamma-Gamma distribution, its probability density function (PDF) is given as
	\begin{align}
		f_{I_a}(x)=\frac{2(\alpha\beta)^{(\alpha+\beta)/2}}{\Gamma(\alpha)\Gamma(\beta)}x^{(\alpha+\beta)/2-1}\text{K}_{\alpha-\beta}\Big(2\sqrt{\alpha\beta x}\Big),
	\end{align}
	where $\text{K}_v(\cdot)$ represents the $v^{th}$ order modified Bessel function of second kind, and $\alpha$ and $\beta$ are the fading coefficients of the atmospheric turbulence.
	Further, the pointing error can be modeled as 
	\begin{align}\label{rad}
		f_{I_p}(x)=\frac{\xi^2}{A_0^{\xi^2}}x^{\xi^2-1},~~~~ \text{for } 0\leq x \leq A_0
	\end{align}
	where $\xi$ is the ratio of the equivalent beam radius to the standard deviation of jitter at the receiver. Further, $A_0$ represents the fraction of the receiver power at $r=0$, where $r$ is the radial displacement at the receiver. 
	
	As the FSO link is assumed to follow the Gamma-Gamma distribution with pointing error impairments, the PDF of its SNR ($\gamma_R$) for both the IM/DD and heterodyne detection techniques at the FSO receiver is given as
	{
		\begin{align}\label{GammaPDF}
			f_{\gamma_R}(x)&=\frac{\xi^2}{i\Gamma(\alpha)\Gamma(\beta)x}\text{G}^{3,0}_{1,3}\Big[\alpha\beta\Big(\frac{x}{\mu_i}\Big)^{1/i}\Big{|}^{\xi^2+1}_{\xi^2,\alpha,\beta}\Big].
		\end{align}
	}
	{Now (\ref{GammaPDF}) is substituted in $F_{{\gamma}_{R}}(x)=\int_{0}^{x}f_{{\gamma}_{R}}(\gamma)d\gamma$ and  (\ref{complex}) \cite[(07.34.21.0084.01)]{wolframe} is used  as} 
		{\begin{gather}
				\int_0^a x^{c_1-1}\text{G}^{m,n}_{p,q}\Big[c_2x^{l/i}\Big{|}^{a_1,...,a_n,a_{n+1},...,a_p}_{b_1,...,b_m,b_{m+1},...,b_q}\Big]dx=\frac{i^{a^*}}{(2\pi)^{c^*(i-1)}a^{-c_1}}\nonumber\\
				\text{G}^{mi,ni+l}_{pi+l,qi+l}\Big[\frac{c_2^ia^l}{i^{i(q-p)}}\Big{|}^{\frac{1-c_1}{l},...,\frac{l-c_1}{l}, \frac{a_1}{i},...,\frac{a_1+i-1}{i},...,\frac{a_n}{i},...,\frac{a_n+i-1}{i},\frac{a_{n+1}}{i},...,\frac{a_{n+1}+i-1}{i},...,\frac{a_p}{i},...,\frac{a_p+i-1}{i}}_{\frac{b_1}{i},...,\frac{b_1+i-1}{i},...,\frac{b_m}{i},...,\frac{b_m+i-1}{i},\frac{b_{m+1}}{i},...,\frac{b_{m+1}+i-1}{i},...,\frac{b_q}{i},...,\frac{b_q+i-1}{i},\frac{-c_1}{l},...,\frac{l-c_1-1}{l}}\Big],
				\label{complex}	
		\end{gather}}
	{where $c^*=m+n-\frac{p+q}{2}$, and $a^*=\sum_{j=1}^{q}b_j-\sum_{j=1}^{p}a_j+\frac{p-q}{2}+1$.
		Solving (\ref{complex}), the cumulative distribution function (CDF) of the SNR of the FSO link for both the detection techniques is given as }
	\begin{align}\label{Gamma_CDF1}
		F_{\gamma_R}(x)=A\text{G}^{3i,1}_{i+1,3i+1}\Big[\frac{B}{\mu_i}x\Big{|}^{1, \tau_1}_{\tau_2, 0}\Big],
	\end{align}
	where $A=\frac{i^{(\alpha+\beta-2)}\xi^2}{(2\pi)^{i-1}\Gamma(\alpha)\Gamma(\beta)}$, $B=\frac{(\alpha \beta)^i}{i^{2i}}$, and $\mu_i$ represents the average SNR of the FSO link for both the detection techniques. For heterodyne detection, $i=1$, and hence, $\mu_1=\bar{\gamma}_R$. For the IM/DD detection, $i=2$, and hence, $\mu_2=\frac{\xi^2\alpha\beta(\xi^2+2)}{(\alpha+1)(\beta+1)(\xi^2+1)^2}\bar{\gamma}_R$.
	Further, $\tau_1=\big[\frac{\xi^2+1}{i},...,\frac{\xi^2+i}{i}\big]$ and $\tau_2=\Big[\frac{\xi^2}{i},...,\frac{\xi^2+i-1}{i},\frac{\alpha}{i},...,\frac{\alpha+i-1}{i},\frac{\beta}{i},...,\frac{\beta+i-1}{i}\Big]$.
	
	The UAV is equipped with $N_t$ antennas  and single antenna is considered at each of the N terrestrial UEs. Further, UAV to the $n^{th}$ terrestrial UE link is assumed to follow Nakagami-m distribution with fading severity $m$ and average power $\Omega$. Thus, the  PDF and CDF of the SNR at the $n^{th}$ UE are given as \cite{tang2005transmit}
	\begin{align}\label{CPDF_Nak}
		f_{\gamma_{RU_n}}(x)&=\frac{1}{\Gamma(mN_t)}\Big(\frac{m}{\bar{\gamma}_{U}}\Big)^{mN_t}x^{mN_t-1}e^{-\frac{m}{\bar{\gamma}_{U}}x}\nonumber\\
		F_{\gamma_{RU_n}}(x)&=1-\frac{1}{\Gamma(mN_t)}\Gamma\Big(mN_t,\frac{m}{\bar{\gamma}_{U}}x\Big),
	\end{align}
	respectively, where $\bar{\gamma}_{U}=\Omega\bar{\gamma}_{RU_n}$ and $\Gamma(\cdot)$ is the complete gamma function. It is assumed that the UEs are in closed proximity and following the independent and identical distribution (i.i.d.). 
	For multiuser diversity, opportunistic scheduling of the UEs is performed.  Thus, the channel quality between the UAV and the $n^{th}$ UE is estimated for the transmission. For this, first the strongest link between the UAV and the set of UEs is selected by the UAV and the index of the selected UE is feedback to the satellite. Therefore, after opportunistic scheduling, the instantaneous SNR of the UAV to terrestrial UE link is given as
	$\gamma_{RU}=\underset{\overset{n=1,2,...,N}{}}{\text{Max}}\gamma_{RU_n}$.
	Applying the order statistics, the PDF of $\gamma_{RU}$ is given as
	\begin{align}\label{order}
		f_{\gamma_{RU}}(x)= N (F_{\gamma_{RU_n}}(x))^{N-1}f_{\gamma_{RU_n}}(x).
	\end{align}
	
	Substituting the PDF and CDF of the SNR of UAV to $n^{th}$ UE link from (\ref{CPDF_Nak}) in (\ref{order}), utilizing the series form of the upper incomplete Gamma function, and expanding the multinomial using \cite[(0.314)]{gradshteyn2000table}, the PDF of $\gamma_{RU}$ is given as
	\begin{align}\label{GRU}
		f_{\gamma_{RU}}(x)&=N\sum_{k=0}^{N-1}\binom{N-1}{k}(-1)^k\sum_{l=0}^{k(mN_t-1)}\varphi_l^k\Big(\frac{m}{\bar{\gamma}_{U}}\Big)^{mN_t+l}
		\frac{1}{\Gamma(mN_t)}x^{mN_t+l-1}e^{-\frac{m}{\bar{\gamma}_{U}}(k+1)x},
	\end{align}
	where 	
	$\varphi_{l}^k$ is recursively calculated as 
	$\varphi_{0}^k=(\delta_0)^k$, $\varphi_{1}^k=k(\delta_1)$, 
	$\varphi_{l}^k=\frac{1}{l\delta_0}\sum_{q=1}^{l}\Big[(qk-l+q)\delta_q\varphi^k_{l-q}\Big]$ for $2\leq l\leq(mN_t-1)$,
	$\varphi_{l}^k=\frac{1}{l\delta_0}\sum_{q=1}^{mN_t-1}\Big[(qk-l+q)\delta_q\varphi^k_{l-q}\Big]$ for $mN_t\leq l< k(mN_t-1)$, and $\varphi_{k(mN_t-1)}^k=(\delta_{mN_t-1})^k$, wherein $\delta_{l}=\frac{1}{l!}$ \cite{kumar2017aser}.
	
	In practice, the channel statistics changes due to its time varying nature and hence, outdated CSI is obtained at the UAV. Delay occurs during the user selection and signal transmission from the UAV. Let, $\hat{\gamma}_{RU}$ is the delayed version of the ${\gamma}_{RU}$ since $\hat{\gamma}_{RU_n}$ is the delayed version of the ${\gamma}_{RU_n}$. Let, $\hat{\gamma}_{RU_n}$ is correlated to ${\gamma}_{RU_n}$ with $\rho=J_0(2\pi f_d \tau)$, where $f_d$ is the Doppler frequency, $\tau$ is the time delay, and $J_0(\cdot)$ is the zeroth order Bessel function of first kind. According to the order statistics, $\hat{\gamma}_{RU}$ will be the induced order statistics of ${\gamma}_{RU}$ \cite{tang2005transmit}. Therefore, the PDF of $\hat{\gamma}_{RU}$ can be given as
	\begin{align}\label{outd}
		f_{\hat{\gamma}_{RU}}(x)= \int_{0}^{\infty}f_{\hat{\gamma}_{RU}|\gamma_{RU}}(x|y)f_{{\gamma}_{RU}}(y) dy,
	\end{align}
	{where $f_{\hat{\gamma}_{RU}|\gamma_{RU}}(x|y)$ is the conditional probability of $\hat{\gamma}_{RU}$ given ${\gamma}_{RU}$.} As $\hat{\gamma}_{RU}$ and ${\gamma}_{RU}$ are the correlated Gamma distributed random variables, their conditional probability is given as
	\begin{align}\label{cond}
		f_{\hat{\gamma}_{RU}|\gamma_{RU}}(x|y)&=\frac{1}{1-\rho}\Big(\frac{m}{\bar{\gamma}_{U}}\Big)\Big(\frac{x}{\rho y}\Big)^{\frac{(mN_t-1)}{2}}e^{-\frac{m(x+\rho y)}{\bar{\gamma}_{U}(1-\rho)}} {I}_{mN_t-1}\Big(\frac{2m\sqrt{\rho x y}}{\bar{\gamma}_{U} (1-\rho)}\Big),
	\end{align}
	where $I_v(\cdot)$ represents the $v^{th}$ order modified Bessel function of first kind. Substituting the values from (\ref{GRU}) and (\ref{cond}) in (\ref{outd}), considering the series expansion of $I_v(\cdot)$, and after some mathematical computations,
	we get
	\begin{align}\label{outdF}
		f_{\hat{\gamma}_{RU}}(x)&=N\sum_{k=0}^{N-1}\binom{N-1}{k}\frac{(-1)^k}{\Gamma(mN_t)}\sum_{l=0}^{k(mN_t-1)}\varphi_l^k
		\sum_{j=0}^{l}\binom{l}{j}\Big(\frac{m}{\bar{{\gamma}}_{U}}\Big)^{(mN_t+j)}\frac{\Gamma(mN_t+l)}{\Gamma(mN_t+j)}\nonumber\\&
		\times\Big(\frac{\rho^j(1-\rho)^{l-j}}{(1+k(1-\rho))^{(mN_t+l+j)}}\Big) x^{mN_t+j-1}e^{-\big(\frac{m(k+1)x}{\bar{{\gamma}}_{U}(1+k(1-\rho))}\big)}.
	\end{align}
	
	Substituting (\ref{outdF}) in $F_{\hat{\gamma}_{RU}}(x)=\int_{0}^{x}f_{\hat{\gamma}_{RU}}(\gamma)d\gamma$, the CDF of the outdated SNR of the UAV to selected terrestrial UE link is given as 
	\begin{align}\label{outdF_CDF}
		F_{\hat{\gamma}_{RU}}(x)&=N\underset{\overset{k,l,j}{}}{\sum}\mathbb{C}_0\mathbb{C}_2^{-\mathbb{C}_1}\Upsilon\Big(\mathbb{C}_1, \mathbb{C}_2x\Big),
	\end{align}
	where $\Upsilon(\cdot,\cdot)$ represents the lower incomplete Gamma function,
	$\underset{\overset{k,l,j}{}}{\sum}=\sum_{k=0}^{N-1}\sum_{l=0}^{k(mN_t-1)}\sum_{j=0}^{l}$, $\mathbb{C}_0=\binom{N-1}{k}\binom{l}{j}\frac{(-1)^k}{\Gamma(mN_t)}\varphi_l^k\Big(\frac{m}{\bar{{\gamma}}_{U}}\Big)^{mN_t+j}\frac{\rho^j(1-\rho)^{l-j}\Gamma(mN_t+l)}{(1+k(1-\rho))^{(mN_t+l+j)}\Gamma(mN_t+j)}$, $\mathbb{C}_1=mN_t+j$, and $\mathbb{C}_2=\frac{m(k+1)}{\bar{{\gamma}}_{U}(1+k(1-\rho))}$.

	\section{Outage Performance}
	\subsection{Outage Probability}
	Outage probability is defined as the probability that the e2e instantaneous SNR reaches below a predefined threshold level ($\gamma_{th}$). Selective DF protocol is used at the UAV. After applying the order statistics to select the strongest user and considering the outdated CSI, the e2e instantaneous received SNR can be expressed as
	$\gamma_{e2e}=\phi(\gamma_R)\hat{\gamma}_{RU}$. Thus, the outage probability is given as
	\begin{align}\label{outage}
		\mathcal{P}_o(\gamma_{th})&=\mathcal{P}[\phi(\gamma_R)\hat{\gamma}_{RU} < \gamma_{th}],\nonumber\\&
		=\mathcal{P}[\phi(\gamma_R)=0]+ \mathcal{P}[\phi(\gamma_R)=1]\mathcal{P}[\hat{\gamma}_{RU} < \gamma_{th}],
	\end{align}
	\vspace{-1em}
	where,
	\begin{align}\label{out_SR}
		\mathcal{P}[\phi(\gamma_R)=0]&=\mathcal{P}[\phi(\gamma_R)<\delta_{th}]=F_{\gamma_R}(\delta_{th}),\nonumber\\ 
		\mathcal{P}[\phi(\gamma_R)=1]&=\mathcal{P}[\phi(\gamma_R)\geq\delta_{th}]=1-F_{\gamma_R}(\delta_{th}).
	\end{align}
	
	Further,  $\mathcal{P}[\hat{\gamma}_{RU} < \gamma_{th}]=F_{\hat{\gamma}_{RU}}(\gamma_{th})$ which is given in (\ref{outdF_CDF}).
	Finally, substituting the values from (\ref{outdF_CDF}) and (\ref{out_SR}) in (\ref{outage}), the outage probability for the considered system is given as
	\begin{align}\label{OutF}
		\mathcal{P}_o(\gamma_{th})&=F_{\gamma_R}(\delta_{th})+(1-F_{\gamma_R}(\delta_{th}))F_{\hat{\gamma}_{RU}}(\gamma_{th}),\nonumber\\&
		=A\text{G}^{3i,1}_{i+1,3i+1}\Big[\frac{B\delta_{th}}{\mu_i}\Big{|}^{1, \tau_1}_{\tau_2, 0}\Big]+N\underset{\overset{k,l,j}{}}{\sum}\mathbb{C}_0\mathbb{C}_2^{-\mathbb{C}_1}\Big(1-A\text{G}^{3i,1}_{i+1,3i+1}\Big[\frac{B\delta_{th}}{\mu_i}\Big{|}^{1, \tau_1}_{\tau_2, 0}\Big]\Big)\Upsilon(\mathbb{C}_1, \mathbb{C}_2\gamma_{th}).
	\end{align}
	\vspace{-1em}
	{
		\subsection{Diversity Order}
		The outage probability obtained in (\ref{OutF}) highlights various insights about the performance of the considered system. However, obtaining the diversity order from  (\ref{OutF}) is quite tricky due to its complex nature. Therefore, in this subsection, an asymptotic outage probability expression is derived by performing the high SNR approximation on the outage probability. For this, the transmit SNR tends to infinity, and hence, $\gamma_R,\gamma_U \rightarrow \infty$.  Thus, the outage probability (\ref{OutF}) can be approximated as 
		\begin{align}\label{OutFApp}
			\mathcal{P}_o^{Asym}(\gamma_{th})\approx F_{\gamma_R}(\delta_{th})+F_{\hat{\gamma}_{RU}}(\gamma_{th}).
		\end{align}
		It is to be noted that the product term from (\ref{OutF}) is neglected at high SNR as it tends to zero.}
	
	{\textit{Corollary 1:} The approximate expression of the asymptotic outage probability can be given as
		\begin{align}\label{Out_Asym}
			\mathcal{P}_o^{Asym}(\gamma_{th})&\approx
			A\sum_{p=1}^{3i}\Big(\frac{B\delta_{th}}{\mu_i}\Big)^{\tau_4,p}	\frac{\prod_{\underset{q\neq p}{q=1}}^{3i} \Gamma(\tau_{4,q}-\tau_{4,p})\prod_{\underset{q=1}{}}^{1}\Gamma(1-\tau_{3,q}+\tau_{4,p})}{\prod_{\underset{q=2}{}}^{i+1}\Gamma(\tau_{3,q}-\tau_{4,p})\prod_{\underset{3i+1}{}}^{3i+1}\Gamma(1-\tau_{4,q}+\tau_{4,p})}\nonumber\\&
			+\begin{cases}
				F_{\hat{\gamma}_{RU1}}(\gamma_{th}) \text{ for $\rho = 1$} \\
				F_{\hat{\gamma}_{RU2}}(\gamma_{th}) \text{ for $\rho < 1$}
			\end{cases}
		\end{align} 
		Here, $F_{\hat{\gamma}_{RU1}}(\gamma_{th})$ represents the approximate high SNR CDF expression of the terrestrial link for the perfect CSI case  which is given as 
		\begin{align}\label{RF_per}
			\vspace{-1em}
			F_{\hat{\gamma}_{RU1}}(\gamma_{th}) \approx\Big(\frac{1}{\Gamma(mN_t+1)}\Big)^{N}\Big(\frac{m \gamma_{th}}{\bar{\gamma}_U}\Big)^{NmN_t}.
		\end{align}
		Further, $F_{\hat{\gamma}_{RU2}}(\gamma_{th})$ represents the approximate high SNR CDF expression of the terrestrial link in case of outdated CSI which is given as
		\begin{align}\label{RF_out}
			F_{\hat{\gamma}_{RU2}}(\gamma_{th}) \approx N\underset{\overset{k,l}{}}{\sum}~\mathbb{D}_0\Big(\frac{m \gamma_{th}}{\bar{\gamma}_U}\Big)^{mN_t},
		\end{align}
		where 
		$\underset{\overset{k,l}{}}{\sum}=\sum_{k=0}^{N-1}\sum_{l=0}^{k(mN_t-1)}$ and $\mathbb{D}_0=\binom{N-1}{k}\frac{(-1)^k}{(\Gamma(mN_t))^2}\varphi_l^k\frac{\Gamma(mN_t+l)}{mN_t}(1+k(1-\rho))^{-(mN_t+l)}(1-\rho)^l.
		\vspace{1em}$
		Proof: See Appendix.}
	
	{From (\ref{FSOapp}), it is observed that the dominant term of meijer-G function is $\min(\frac{\xi^2}{i},\frac{\alpha}{i},\frac{\beta}{i})$ (defined by the smallest negative power of the transmit SNR term (consequently $\gamma_R$)).}
	
	{Case-1: For the perfect CSI case, using (\ref{RF_per}) in (\ref{Out_Asym}), the diversity order for the considered system can be obtained as $\min(\frac{\xi^2}{i},\frac{\alpha}{i},\frac{\beta}{i}, NmN_t)$.} 
	
	{Case-2: For the outdated CSI case, using (\ref{RF_out}) in (\ref{Out_Asym}), the diversity order for the considered system can be obtained as $\min(\frac{\xi^2}{i},\frac{\alpha}{i},\frac{\beta}{i}, mN_t)$.} 
	
	{From the above, it can be concluded that the diversity order depends upon the atmospheric turbulence, pointing error, FSO detection type, number of users, fading parameter, and number of transmit antennas for the perfect CSI case, however,
		the multiuser diversity effect is vanished in case of outdated CSI.}
	\section{Capacity Analysis}
	In this section, we will derive the ergodic and the effective capacity (in bits/s/Hz) of the considered network.
	\vspace{-1em}
	\subsection{Ergodic Capacity}
	Statistical expectation of the mutual information between the end users gives us the ergodic capacity (bits/s/Hz) of the considered system which can be expressed as \cite{zedini2020performance,lapidoth2009capacity,alouini1999capacity,huang2019performance}
		\begin{align}\label{Cap}
			\mathcal{C}_{e}&=\frac{1}{2}\mathbb{E}[\text{log}_2(1+\varrho\,\gamma_{e2e})]\nonumber\\
			&=\frac{1}{2}\mathbb{E}[\text{log}_2(1+\varrho\,\phi(\gamma_R)\hat{\gamma}_{RU})]\nonumber\\
			&=\frac{1}{2}{\mathcal{P}[\phi(\gamma_R)=1]}~\underset{\overset{I_1}{}}{\underbrace{\mathbb{E}[\text{log}_2(1+\varrho\,\hat{\gamma}_{RU})]}}.
		\end{align}
		Here, $\varrho=1$ for the heterodyne detection ($i=1$) and   $\varrho=\frac{e}{2\pi}$ for the IM/DD ($i=2$). It is to be noted that (\ref{Cap}) is exact for heterodyne detection ($i=1$) and is the lower-bound for the IM/DD ($i=2$).
	From (\ref{Cap}),  $I_1$ can be solved as
	\begin{align}\label{Cap1}
		I_1=&\int_{0}^{\infty}\text{log}_2(1+\varrho\,\gamma)f_{\hat{\gamma}_{RU}}(\gamma)d\gamma.
	\end{align}
	Invoking (\ref{outdF}) in (\ref{Cap1}), and solving the required integral with the help of \cite[(8.4.6.5))]{prudnikov1990integrals}, we get
	\begin{align}\label{Cap2}
		I_1=&\frac{N}{\text{ln}(2)}\underset{\overset{k,l,j}{}}{\sum}\mathbb{C}_0 \mathbb{C}_2^{-\mathbb{C}_1}\text{G}^{1,3}_{3,2}\Big[\frac{\varrho}{\mathbb{C}_2}\Big{|}^{-\mathbb{C}_1+1,1,1}_{1, 0}\Big].
	\end{align}
	Substituting the identities from (\ref{out_SR}) and (\ref{Cap2}) in (\ref{Cap}),  final expression of ergodic capacity is derived as
	\begin{align}\label{CapF}
		\mathcal{C}_{e}&=\frac{N}{2\,\text{ln}(2)}\underset{\overset{k,l,j}{}}{\sum}\mathbb{C}_0 \mathbb{C}_2^{-\mathbb{C}_1}\text{G}^{1,3}_{3,2}\Big[\frac{\varrho}{\mathbb{C}_2}\Big{|}^{-\mathbb{C}_1+1,1,1}_{1, 0}\Big]\Big(1-A\text{G}^{3i,1}_{i+1,3i+1}\Big[\frac{B\delta_{th}}{\mu_i}\Big{|}^{1, \tau_1}_{\tau_2, 0}\Big]\Big).
	\end{align}
	\vspace{-1em}
	\subsection{Effective Capacity}
	{The emerging next generation wireless communication  applications such as multimedia streaming, interactive gaming, autonomous vehicles, mobile video telephony, etc require low latency. 
		The extensively used ergodic capacity metric is unable to measure the quality of service (QoS) of such emerging real-time applications.
		Hence, effective capacity, a QoS aware link-layer channel model is used in the wireless networks for performance analysis under such delay constraints \cite{zhang2015effective}.}
	Effective capacity guarantees system's minimum delay, and maintains an efficient constant data-rate. Higher value of effective capacity guarantees higher QoS and lower delay of the considered system. Effective capacity can be expressed as
	\begin{align}\label{Ecap}
		\mathcal{C}_{eff}(\Theta)=-\frac{1}{\Theta}\text{ln}(\mathbb{E}[e^{-\Theta r_{e2e}}]),
	\end{align} 
	where $r_{e2e}$ is the target rate of the communication system and $\Theta$ is the quality parameter. Increase in $\Theta$ limits the effective capacity of the communication system.
	The target rate (in  bits/s/Hz) of the considered communication system can be given as $r_{e2e}=\frac{1}{2}\text{log}_2(1+\varrho\,\gamma_{e2e})$.
	Substituting $r_{e2e}$ in (\ref{Ecap}), we obtain
	\begin{align}
		\mathcal{C}_{eff}(\Theta)=-\frac{1}{\Theta}\text{ln}\Big(\mathbb{E}[e^{-\frac{\Theta}{2\text{ln}(2)}\; \text{ln}(1+\varrho\,\gamma_{e2e})}]\Big).
	\end{align} 
	After mathematical simplification, we get
	\begin{align} \label{Eff1}
		\mathcal{C}_{eff}(\Theta)
		=-\frac{1}{\Theta}\text{ln}\Big(\mathbb{E}[(1+\varrho\,\gamma_{e2e})^{-\hat{\Theta}}]\Big),
	\end{align}
	where $\hat{\Theta}=\frac{\Theta}{2\text{ln}(2)}$.
	
	Substituting the $\gamma_{e2e}$ in (\ref{Eff1}), $\mathcal{C}_{eff}(\Theta)$ can be written as 
	
	\begin{align} \label{Eff2}
		\mathcal{C}_{eff}(\Theta)&
		=-\frac{1}{\Theta}\text{ln}\Big(\mathbb{E}[(1+\varrho\,\phi(\gamma_R)\hat{\gamma}_{RU})^{-\hat{\Theta}}]\Big).
	\end{align}
	Further, (\ref{Eff2})  can be written as
	\begin{align} \label{Eff3}
		\mathcal{C}_{eff}(\Theta)&
		=-\frac{1}{\Theta}{\mathcal{P}[\phi(\gamma_R)=1]}~\text{ln}\Big(\mathbb{E}[(1+\varrho\,\hat{\gamma}_{RU})^{-\hat{\Theta}}]\Big),\nonumber\\&
		=-\frac{1}{\Theta}{\mathcal{P}[\phi(\gamma_R)=1]}~\text{ln}\Big(\int_{0}^{\infty}(1+\varrho\,\gamma)^{-\hat{\Theta}}f_{\hat{\gamma}_{RU})}(\gamma)d\gamma \Big).
	\end{align}
	Substituting $f_{\hat{\gamma}_{RU})}(\gamma)$ from (\ref{outdF}) in (\ref{Eff3}), effective capacity in integral form can be expressed as
	\begin{align} \label{Eff4}
		\mathcal{C}_{eff}(\Theta)&
		=-\frac{1}{\Theta}{\mathcal{P}[\phi(\gamma_R)=1]}\text{ln}\Big(N\underset{\overset{k,l,j}{}}{\sum}\mathbb{C}_0\int_{0}^{\infty}(1+\varrho\,\gamma)^{-\hat{\Theta}}~\gamma^{\mathbb{C}_1-1}e^{-\mathbb{C}_2\gamma}d\gamma \Big).
	\end{align}
	After change of variables and with some mathematical computations, we get
	\begin{align} 
		\mathcal{C}_{eff}(\Theta)&
		=-\frac{1}{\Theta}{\mathcal{P}[\phi(\gamma_R)=1]}~
		\text{ln}\Big(N\underset{\overset{k,l,j}{}}{\sum}\mathbb{C}_0e^{\mathbb{C}_2}\sum_{z1=0}^{\mathbb{C}_1-1}\binom{\mathbb{C}_1-1}{z1}(-1)^{\mathbb{C}_1-1-z1}\int_{1}^{\infty}t^{z1-\hat{\Theta}}~e^{-\mathbb{C}_2 t}dt \Big).
	\end{align}
	Solving the required integral with the help of \cite[(3.351)]{gradshteyn2000table}, analytical expression of effective capacity can be derived as
	\begin{align} \label{Eff_F}
		\mathcal{C}_{eff}(\Theta)&
		=-\frac{1}{\Theta}\Big(1-A\text{G}^{3i,1}_{i+1,3i+1}\Big[\frac{B\delta_{th}}{\mu_i}\Big{|}^{1, \tau_1}_{\tau_2, 0}\Big]\Big)	\text{ln}\Big(N\underset{\overset{k,l,j}{}}{\sum}\mathbb{C}_0\Big(\frac{1}{\varrho}\Big)^{\mathbb{C}_1-z_1+\hat{\Theta}-1}e^{\frac{\mathbb{C}_2}{\varrho}}\sum_{z1=0}^{\mathbb{C}_1-1}\binom{\mathbb{C}_1-1}{z1}\nonumber\\&
		\times
		(-1)^{\mathbb{C}_1-1-z1}\mathbb{C}_2^{-(z1-\hat{\Theta}+1)}\Gamma\Big((z1-\hat{\Theta}+1),\mathbb{C}_2\Big)\Big).
	\end{align}

	\section{ASER Analysis}
	For any modulation scheme, the CDF based generalized ASER expression for a given instantaneous e2e SNR is given as 
	\vspace{-0.5em}
	\begin{equation}\label{error}
		\mathcal{P}_e=-\int_{0}^{\infty}\mathcal{P}_s^{'}(e|\gamma)\mathcal{P}_{o}(\gamma)d\gamma,
	\end{equation}	
	where $\mathcal{P}_{o}(\gamma)$ represents the outage probability of the considered system and $\mathcal{P}_s^{'}(e|\gamma)$ is the first order derivative of the conditional SEP \big($\mathcal{P}_s(e|\gamma)$\big) of the received SNR.

	\subsection{Hexagonal QAM}
	For the AWGN channel, generalized conditional SEP expression of $M$-ary HQAM constellation is given as
	\begin{align}\label{CHQAM}
		\mathcal{P}^{H}_s(e|\gamma)&= K Q(\sqrt{\theta\gamma})
		+\frac{2}{3}K_cQ^2\Big(\sqrt{\frac{2\theta\gamma}{3}}\Big)
		-2K_cQ(\sqrt{\theta\gamma})Q\Big(\sqrt{\frac{\theta\gamma}{3}}\Big),
	\end{align}
	where $\theta$, $K$, and $K_c$ are the constants and their values for various irregular HQAM (optimum) constellations are given in Table \ref{tab_HQAM}. 
	The first order derivative of \big($\mathcal{P}_s(e|\gamma)$\big) can be given as \cite{kumar2017aser,singya2017impact}
	
	\begin{align}\label{DHQAM}
		\mathcal{P^{'}}_s^{H}(e|\gamma)&=\frac{1}{2}\sqrt{\frac{\theta}{2\pi}}(K_c-K)\gamma^{-\frac{1}{2}}e^{-\frac{\theta\gamma}{2}}
		-\frac{K_c}{3}\sqrt{\frac{\theta}{3\pi}}\gamma^{-\frac{1}{2}}e^{-\frac{\theta\gamma}{3}}
		+\frac{K_c}{2}\sqrt{\frac{\theta}{6\pi}}\gamma^{-\frac{1}{2}}e^{-\frac{\theta\gamma}{6}}\nonumber\\&
		+\frac{2K_c\theta}{9\pi}e^{-\frac{2\theta}{3}\gamma}\;_1F_1\Big(1;\frac{3}{2};\frac{\theta}{3}\gamma\Big)
		-\frac{K_c\theta e^{-\frac{2\theta}{3}\gamma}}{2\sqrt{3}\pi}
		\Big[\;_1F_1\Big(1;\frac{3}{2};\frac{\theta}{2}\gamma\Big)
		+\;_1F_1\Big(1;\frac{3}{2};\frac{\theta}{6}\gamma\Big)\Big].
	\end{align}
	
	Invoking $\mathcal{P}_{o}(\gamma)$ from (\ref{OutF}) and   $\mathcal{P^{'}}_s^{H}(e|\gamma)$ from (\ref{DHQAM}) into (\ref{error}),  and using the identities from \cite[(6.455), (7.813)]{gradshteyn2000table} to solve the required integrals, the generalized ASER expression of HQAM constellation is derived as 
		\begin{align}\label{HQAMF}
			\mathcal{P}^{H}&=\frac{1}{2}\sqrt{\frac{\theta}{2\pi}}(K-K_c)\bigg[A\mathbb{F}\Big(\frac{-1}{2},\frac{\theta}{2}\Big)+N\underset{\overset{k,l,j}{}}{\sum}\mathbb{C}_0\mathbb{C}_2^{-\mathbb{C}_1}\bigg\{\mathbb{G}\Big(\frac{1}{2},\frac{\theta}{2}\Big)-A\Gamma(\mathbb{C}_1)\bigg(\mathbb{F}\Big(\frac{-1}{2},\frac{\theta}{2}\Big)
			\nonumber\\&
			+\sum_{z=0}^{\mathbb{C}_1-1}\frac{\mathbb{C}_2^z}{z!}\mathbb{F}\Big(-z-\frac{1}{2},\mathbb{C}_2+\frac{\theta}{2}\Big)\bigg)\bigg\}\bigg]
			+\frac{K_c}{3}\sqrt{\frac{\theta}{3\pi}}\bigg[A\mathbb{F}\Big(\frac{-1}{2},\frac{\theta}{3}\Big)+N\underset{\overset{k,l,j}{}}{\sum}\mathbb{C}_0\mathbb{C}_2^{-\mathbb{C}_1}\bigg\{\mathbb{G}\Big(\frac{1}{2},\frac{\theta}{3}\Big)
			\nonumber\\&
			-A\Gamma(\mathbb{C}_1)\bigg(\mathbb{F}\Big(\frac{-1}{2},\frac{\theta}{3}\Big)+\sum_{z=0}^{\mathbb{C}_1-1}\frac{\mathbb{C}_2^z}{z!}\mathbb{F}\Big(-z-\frac{1}{2},\mathbb{C}_2+\frac{\theta}{3}\Big)\bigg)\bigg\}\bigg]-\frac{K_c}{2}\sqrt{\frac{\theta}{6\pi}}\bigg[A\mathbb{F}\Big(\frac{-1}{2},\frac{\theta}{6}\Big)\nonumber\\&
			+N\underset{\overset{k,l,j}{}}{\sum}\mathbb{C}_0\mathbb{C}_2^{-\mathbb{C}_1}\bigg\{\mathbb{G}\Big(\frac{1}{2},\frac{\theta}{6}\Big)-A\Gamma(\mathbb{C}_1)\bigg(\mathbb{F}\Big(\frac{-1}{2},\frac{\theta}{6}\Big)+\sum_{z=0}^{\mathbb{C}_1-1}\frac{\mathbb{C}_2^z}{z!}\mathbb{F}\Big(-z-\frac{1}{2},\mathbb{C}_2+\frac{\theta}{6}\Big)\bigg)\bigg\}\bigg]\nonumber\\&
			-\sum_{z_1=0}^{\infty}\mathbb{C}_3\bigg[A\mathbb{F}\Big({-z_1-1},\frac{2\theta}{3}\Big)+N\underset{\overset{k,l,j}{}}{\sum}\mathbb{C}_0\mathbb{C}_2^{-\mathbb{C}_1}\bigg\{\mathbb{G}\Big(z_1+1,\frac{2\theta}{3}\Big)-A\Gamma(\mathbb{C}_1)\bigg(\mathbb{F}\Big(-z_1-1,\frac{2\theta}{3}\Big)\nonumber\\&
			+\sum_{z=0}^{\mathbb{C}_1-1}\frac{\mathbb{C}_2^z}{z!}\mathbb{F}\Big(-z-z_1-1,\mathbb{C}_2+\frac{2\theta}{3}\Big)\bigg)\bigg\}\bigg],
		\end{align}
	where $\mathbb{C}_3=\frac{(1)_{z_1}}{(1.5)_{z_1}z_1!}\Big(\frac{2K_c\theta}{9\pi}\Big(\frac{\theta}{3}\Big)^{z_1}-\frac{K_c\theta}{2\sqrt{3}\pi}\Big(\Big(\frac{\theta}{2}\Big)^{z_1}+\Big(\frac{\theta}{6}\Big)^{z_1}\Big)\Big)$,
	$\mathbb{F}(\psi_1,\psi_2)=(\psi_2)^{\psi_1}\text{G}^{3i,2}_{i+2,3i+1}\Big[\frac{B}{\psi_2\mu_i}\Big{|}^{\psi_1+1, 1, \tau_1}_{\tau_2, 0}\Big]$, and
	$\mathbb{G}(\psi_1,\psi_2)=\frac{\mathbb{C}_2^{\mathbb{C}_1}\Gamma(\mathbb{C}_1+\psi_1)}{\mathbb{C}_1(\mathbb{C}_2+\psi_2)^{(\mathbb{C}_1+\psi_1)}}\;_2F_1(1,\mathbb{C}_1+\psi_1, \mathbb{C}_1+1, \frac{\mathbb{C}_2}{\mathbb{C}_2+\psi_2})$.
	
{\renewcommand{\arraystretch}{1.3}
		\begin{table}[!h]
			\centering
			\vspace{-0.5em}
			\caption {{Various selection parameters for  irregular HQAM (optimum) constellations \cite{singya2020performance}.}}
			\label{tab_HQAM}
			\begin{tabular}{|c|ccc||}
				\hline
				&     \multicolumn{3}{c|}{Irregular HQAM (optimum)}       \\ \hline
				M 	& \multicolumn{1}{c|}{$\theta$} & 
				\multicolumn{1}{c|}{$K$}   & \multicolumn{1}{c|}{$K_c$} \\ \hline
				
				4   &  \multicolumn{1}{c|}{1} & \multicolumn{1}{c|}{$\frac{5}{2}$} & \multicolumn{1}{c|}{$\frac{3}{2}$} 
				\\ \hline
				
				8    &   \multicolumn{1}{c|}{$\frac{32}{69}$} & \multicolumn{1}{c|}{$\frac{7}{2}$} & \multicolumn{1}{c|}{$\frac{21}{8}$} 
				\\ \hline
				
				16	&  \multicolumn{1}{c|}{$\frac{8}{35}$} & \multicolumn{1}{c|}{$\frac{33}{8}$} & \multicolumn{1}{c|}{$\frac{27}{8}$} 
				\\ \hline
				
				32  &  \multicolumn{1}{c|}{$\frac{512}{4503}$} & \multicolumn{1}{c|}{$\frac{75}{16}$} & \multicolumn{1}{c|}{$\frac{33}{8}$} 
				\\ \hline
				
				64  &  \multicolumn{1}{c|}{$\frac{8}{141}$} & \multicolumn{1}{c|}{$\frac{163}{32}$} & \multicolumn{1}{c|}{$\frac{75}{16}$} 
				\\ \hline
				
				128	 &  \multicolumn{1}{l|}{$\frac{2}{70.56}$} & \multicolumn{1}{l|}{$\frac{343}{64}$}  & \multicolumn{1}{c|}{$\frac{81}{16}$}
				\\ \hline
				
				256  &    \multicolumn{1}{c|}{$\frac{2}{141}$}   & \multicolumn{1}{c|}{$\frac{711}{128}$} & \multicolumn{1}{c|}{$\frac{171}{32}$} 
				\\ \hline
				
				512	 &    \multicolumn{1}{c|}{$\frac{200}{28217}$} & \multicolumn{1}{c|}{$\frac{2911}{512}$} & \multicolumn{1}{c|}{$\frac{5667}{1024}$} 
				\\ \hline
				
				1024   &    \multicolumn{1}{c|}{$\frac{100}{28227}$} & \multicolumn{1}{c|}{$\frac{2955}{512}$} & \multicolumn{1}{c|}{$\frac{1449}{256}$} 
				\\ \hline
			\end{tabular}
		\end{table}
	}
	\subsection{RQAM}
	For the AWGN channel, the conditional SEP  expression for  $M_i\times N_q$-QAM constellation is given in
	\begin{align}\label{CRQAM}
		\mathcal{P}^{R}_s(e|\gamma)=
		2\Big[p_0Q(a_0\sqrt{\gamma})(1-2q_0Q(b_0\sqrt{\gamma}))
		+q_0Q(b_0\sqrt{\gamma})\Big],	
	\end{align}	
	where $p_0=1-\frac{1}{M_i}$,  $q_0=1-\frac{1}{N_q}$, $a_0=\sqrt{\frac{6}{(M_i^2-1)+(N_q^2-1)\beta_R}}$, and $b_0=\beta_Ra_0$. Here $\beta_R=d_Q/d_I$ is the ratio of the quadrature and in-phase decision distances and $M_i$ and $N_q$ are the constellation points of in-phase and quadrature phase, respectively.
	The first order derivative of (\ref{CRQAM}) can be obtained as
	\begin{align}\label{DRQAM}
		\mathcal{P^{'}}^{R}_s(e|\gamma)&=\gamma^{-\frac{1}{2}}\Big[\frac{a_0p_0(q_0-1)}{\sqrt{2\pi}}e^{-\frac{a_0^2\gamma}{2}}
		+\frac{b_0(p_0-1)q_0}{\sqrt{2\pi}}e^{-\frac{b_0^2\gamma}{2}}\Big]\nonumber\\&
		-\frac{a_0b_0p_0q_0}{\pi}e^{-\frac{(a_0^2+b_0^2)\gamma}{2}}\Big[\,_1F_1\Big(1;\frac{3}{2};\frac{a_0^2\gamma}{2}\Big)
		+\,_1F_1\Big(1;\frac{3}{2};\frac{b_0^2\gamma}{2}\Big)\Big].
	\end{align}
	
	Substituting the respective values of $\mathcal{P^{'}}_s^{R}(e|\gamma)$ and $\mathcal{P}_0(\gamma)$ from (\ref{DRQAM}) and (\ref{OutF}) into (\ref{error}),  solving the required integral with the help of \cite[(6.455), (7.813)]{gradshteyn2000table} with some mathematical computations, the generalized ASER expression of  RQAM constellation is derived as
		\begin{align}\label{RQAMF}
			\mathcal{P}^{R}&=\frac{a_0p_0(1-q_0)}{\sqrt{2\pi}}\bigg[A\mathbb{F}\Big(\frac{-1}{2},\frac{a_0^2}{2}\Big)+N\underset{\overset{k,l,j}{}}{\sum}\mathbb{C}_0\mathbb{C}_2^{-\mathbb{C}_1}\bigg\{\mathbb{G}\Big(\frac{1}{2},\frac{a_0^2}{2}\Big)-A\Gamma(\mathbb{C}_1)\bigg(\mathbb{F}\Big(\frac{-1}{2},\frac{a_0^2}{2}\Big)\nonumber\\&
			+\sum_{z=0}^{\mathbb{C}_1-1}\frac{\mathbb{C}_2^z}{z!}\mathbb{F}\Big(-z-\frac{1}{2},\mathbb{C}_2+\frac{a_0^2}{2}\Big)\bigg)\bigg\}\bigg]
			+\frac{b_0q_0(1-p_0)}{\sqrt{2\pi}}\bigg[A\mathbb{F}\Big(\frac{-1}{2},\frac{b_0^2}{2}\Big)+N\underset{\overset{k,l,j}{}}{\sum}\mathbb{C}_0\mathbb{C}_2^{-\mathbb{C}_1}\bigg\{\mathbb{G}\Big(\frac{1}{2},\frac{b_0^2}{2}\Big)\nonumber\\&
			-A\Gamma(\mathbb{C}_1)\bigg(\mathbb{F}\Big(\frac{-1}{2},\frac{b_0^2}{2}\Big)+\sum_{z=0}^{\mathbb{C}_1-1}\frac{\mathbb{C}_2^z}{z!}\mathbb{F}\Big(-z-\frac{1}{2},\mathbb{C}_2+\frac{b_0^2}{2}\Big)\bigg)\bigg\}\bigg]
			+\sum_{z_1=0}^{\infty}\mathbb{C}_4\bigg[A\mathbb{F}\Big({-z_1-1},\frac{a_0^2+b_0^2}{2}\Big)\nonumber\\&
			+N\underset{\overset{k,l,j}{}}{\sum}\mathbb{C}_0\mathbb{C}_2^{-\mathbb{C}_1}\bigg\{\mathbb{G}\Big(z_1+1,\frac{a_0^2+b_0^2}{2}\Big)-A\Gamma(\mathbb{C}_1)\bigg(\mathbb{F}\Big(-z_1-1,\frac{a_0^2+b_0^2}{2}\Big)\nonumber\\&
			+\sum_{z=0}^{\mathbb{C}_1-1}\frac{\mathbb{C}_2^z}{z!}\mathbb{F}\Big(-z-z_1-1,\mathbb{C}_2+\frac{a_0^2+b_0^2}{2}\Big)\bigg)\bigg\}\bigg],
		\end{align}
	  where $\mathbb{C}_4=\frac{a_0b_0p_0q_0}{\pi}\frac{(1)_{z_1}}{(1.5)_{z_1}z_1!}\Big(\Big(\frac{a_0^2}{2}\Big)^{z_1}+\Big(\frac{b_0^2}{2}\Big)^{z_1}\Big)$.

	\subsection{XQAM Constellations}
	Generalized conditional SEP expression of XQAM constellation for the AWGN channel is given as \cite{sadhwani2017simplified}
	\begin{align}\label{CXQAM}
		\mathcal{P}_s^{X}(e|\gamma)&=N_nQ\Big(\sqrt{\frac{2\gamma}{a_1}}\Big)+\frac{8}{M_iN_q}\Big[\sum_{l_1=1}^{\frac{a_2}{2}-1}Q\Big(2l_1\sqrt{\frac{2\gamma}{a_1}}\Big)
		+Q\Big(a_2\sqrt{\frac{2\gamma}{a_1}}\Big)\nonumber\\&
		-2\sum_{l_1=1}^{\frac{a_2}{2}-1}Q\Big(\sqrt{\frac{2\gamma}{a_1}}\Big)Q\Big(2l_1\sqrt{\frac{2\gamma}{a_1}}\Big)
		-Q\Big(\sqrt{\frac{2\gamma}{a_1}}\Big)Q\Big(a_2\sqrt{\frac{2\gamma}{a_1}}\Big)\Big]-a_3Q^2\Big(\sqrt{\frac{2\gamma}{a_1}}\Big),
	\end{align} 
	where $N_n=4-\frac{2(M_i+N_q)}{M_iN_q}$, $a_1=\frac{2}{3}\Big(\frac{31M_iN_q}{32}-1\Big)$, $a_2=\frac{M_i-N_q}{2}$, and $a_3=4-\frac{4(M_i+N_q)}{M_iN_q}+\frac{8}{M_iN_q}$.
	
	The first order derivative of $\mathcal{P}_s^{X}(e|\gamma)$ is given as \cite{parvez2019aser}
	\begin{align}\label{DXQAM}
		\mathcal{P^{'}}_s^{X}(e|\gamma)&=\frac{1}{2\sqrt{\pi a_1}}\Big(-N_n+\frac{12}{M_1N_q}+a_3\Big)\gamma^{-\frac{1}{2}}e^{-\frac{\gamma}{a_1}}-\frac{a_4}{\sqrt{\pi a_1}}\gamma^{-\frac{1}{2}}e^{-a_2^2\frac{\gamma}{a_1}}\nonumber\\&
		-\frac{16}{M_iN_q}\sum_{l_1=1}^{\frac{a_2}{2}-1}\Big[\frac{l_1}{\pi a_1}e^{-a_5 \gamma}
		\Big\{\,_1F_1\Big(1,\frac{3}{2},\frac{\gamma}{a_1}\Big)
		+\,_1F_1\Big(1,\frac{3}{2},\frac{4l_1^2\gamma}{a_1}\Big)\Big\}\Big]-2\frac{a_4}{\pi a_1}\nonumber\\&
		\times e^{-\frac{(1+a_2^2)}{a_1}\gamma}\Big\{\,_1F_1\Big(1,\frac{3}{2},\frac{\gamma}{a_1}\Big)+\,_1F_1\Big(1,\frac{3}{2},\frac{a_2\gamma}{a_1}\Big)\Big\}
		-\frac{a_3}{\pi a_1}e^{-\frac{2\gamma}{a_1}}\,_1F_1\Big(1,\frac{3}{2},\frac{\gamma}{a_1}\Big),
	\end{align}
	where $a_4=\frac{M_i-N_q}{M_iN_q}$, $a_5=\frac{4l_1^2+1}{a_1}$.
	Substituting $\mathcal{P}_{o}(\gamma)$ and $\mathcal{P^{'}}_s^{X}(e|\gamma)$ from (\ref{OutF}) and (\ref{DXQAM}), respectively in (\ref{error}), solving the necessary integrations and after some mathematical computations, the 
	final generalized SEP expression of XQAM constellation is given as 
		\begin{align}\label{XQAMF}
			\mathcal{P}^{X}&=-a_6\bigg[A\mathbb{F}\Big(\frac{-1}{2},\frac{1}{a_1}\Big)+N\underset{\overset{k,l,j}{}}{\sum}\mathbb{C}_0\mathbb{C}_2^{-\mathbb{C}_1}\bigg\{\mathbb{G}\Big(\frac{1}{2},\frac{1}{a_1}\Big)-A\Gamma(\mathbb{C}_1)\bigg(\mathbb{F}\Big(\frac{-1}{2},\frac{1}{a_1}\Big)\nonumber\\&
			+\sum_{z=0}^{\mathbb{C}_1-1}\frac{\mathbb{C}_2^z}{z!}\mathbb{F}\Big(-z-\frac{1}{2},\mathbb{C}_2+\frac{1}{a_1}\Big)\bigg)\bigg\}\bigg]
			+\frac{a_4}{\sqrt{\pi a_1}}\bigg[A\mathbb{F}\Big(\frac{-1}{2},\frac{a_2^2}{a_1}\Big)+N\underset{\overset{k,l,j}{}}{\sum}\mathbb{C}_0\mathbb{C}_2^{-\mathbb{C}_1}
			\nonumber\\&
			\times\bigg\{\mathbb{G}\Big(\frac{1}{2},\frac{a_2^2}{a_1}\Big)-A\Gamma(\mathbb{C}_1)\bigg(\mathbb{F}\Big(\frac{-1}{2},\frac{a_2^2}{a_1}\Big)+\sum_{z=0}^{\mathbb{C}_1-1}\frac{\mathbb{C}_2^z}{z!}\mathbb{F}\Big(-z-\frac{1}{2},\mathbb{C}_2+\frac{a_2^2}{a_1}\Big)\bigg)\bigg\}\bigg]\nonumber\\&
			+\frac{16}{M_iN_q}\sum_{l_1=1}^{\frac{a_2}{2}-1}\frac{l_1}{\pi a_1}\sum_{z_1=0}^{\infty}\mathbb{C}_5\bigg[A\mathbb{F}\Big({-z_1-1},a_5\Big)+N\underset{\overset{k,l,j}{}}{\sum}\mathbb{C}_0\mathbb{C}_2^{-\mathbb{C}_1}\bigg\{\mathbb{G}\Big(z_1+1,a_5\Big)
			\nonumber\\&
			-A\Gamma(\mathbb{C}_1)\bigg(\mathbb{F}\Big(-z_1-1,a_5\Big)
			+\sum_{z=0}^{\mathbb{C}_1-1}\frac{\mathbb{C}_2^z}{z!}\mathbb{F}\Big(-z-z_1-1,\mathbb{C}_2+a_5\Big)\bigg)\bigg\}\bigg]\nonumber\\&
			+\frac{2a_4}{\pi a_1}\sum_{z_1=0}^{\infty}\mathbb{C}_6\bigg[A\mathbb{F}\Big({-z_1-1},\frac{(1+a_2^2)}{a_1}\Big)
			+N\underset{\overset{k,l,j}{}}{\sum}\mathbb{C}_0\mathbb{C}_2^{-\mathbb{C}_1}\bigg\{\mathbb{G}\Big(z_1+1,\frac{(1+a_2^2)}{a_1}\Big)
			\nonumber\\&
			-A\Gamma(\mathbb{C}_1)\bigg(\mathbb{F}\Big(-z_1-1,\frac{(1+a_2^2)}{a_1}\Big)+\sum_{z=0}^{\mathbb{C}_1-1}\frac{\mathbb{C}_2^z}{z!}\mathbb{F}\Big(-z-z_1-1,\mathbb{C}_2+\frac{(1+a_2^2)}{a_1}\Big)\bigg)\bigg\}\bigg]\nonumber\\&+\frac{a_3}{\pi a_1}\sum_{z_1=0}^{\infty}\mathbb{C}_7\bigg[A\mathbb{F}\Big({-z_1-1},\frac{2}{a_1}\Big)
			+N\underset{\overset{k,l,j}{}}{\sum}\mathbb{C}_0\mathbb{C}_2^{-\mathbb{C}_1}\bigg\{\mathbb{G}\Big(z_1+1,\frac{2}{a_1}\Big)
			\nonumber\\&
			-A\Gamma(\mathbb{C}_1)\bigg(\mathbb{F}\Big(-z_1-1,\frac{2}{a_1}\Big)
			+\sum_{z=0}^{\mathbb{C}_1-1}\frac{\mathbb{C}_2^z}{z!}\mathbb{F}\Big(-z-z_1-1,\mathbb{C}_2+\frac{2}{a_1}\Big)\bigg)\bigg\}\bigg],
		\end{align} 
	where $a_6=\frac{1}{2\sqrt{\pi a_1}}\Big(-N_n+\frac{12}{M_1N_q}+a_3\Big)$, $\mathbb{C}_5=\frac{(1)_{z_1}}{(1.5)_{z_1}z_1!}\Big(\Big(\frac{1}{a_1}\Big)^{z_1}+\Big(\frac{4l_1^2}{a_1}\Big)^{z_1}\Big)$, $\mathbb{C}_6=\frac{(1)_{z_1}}{(1.5)_{z_1}z_1!}\Big(\Big(\frac{1}{a_1}\Big)^{z_1}+\Big(\frac{a_2}{a_1}\Big)^{z_1}\Big)$, and $\mathbb{C}_7=\frac{(1)_{z_1}}{(1.5)_{z_1}z_1!}\Big(\frac{1}{a_1}\Big)^{z_1}$.
	
	\section{Theoretical and Simulation Results}
	{In this Section, the numerical results are validated with the simulation results obtained through the Monte-Carlo simulations. For the simulation results, $10^{7}$ realizations have been performed. For the FSO link, product of two random variables which follow generalized Gamma distributions is used to generate the samples of atmospheric turbulence. The pointing error is generated  by considering Rayleigh distribution for radial displacement (\ref{rad}).
		For the analysis, a 
		low earth orbit (LEO) satellite at an altitude of 705 km is considered and the UAV (HAP) is considered in the stratosphere at 17 km altitude from the earth surface.}
	For the FSO link, parameters are taken from \cite{antonini2006feasibility,kong2020multiuser} which are given in Table \ref{para}.  Parameters for the RF links are also mentioned in Table \ref{para}. As the HAP is considered for UAV, the satellite to HAP FSO link experiences weak atmospheric turbulence, hence, the atmospheric turbulence parameters $\alpha=2.902$ and $\beta=2.51$ are considered. Further, the UAV is a moving relay node and hence, and two different values of pointing error i.e., $\xi=1.1$ for severe case and $\xi=6.7$ for negligible pointing error case are considered. {Electrical to optical conversion coefficient $\eta=1$ is considered. The ASER expressions of HQAM, RQAM, and XQAM ((\ref{HQAMF}), (\ref{RQAMF}), and (\ref{XQAMF}), respectively) consist of one infinite series ``$z_1$" which needs to be truncated to a finite value to get the numerical results from these mathematical expressions. Hence, it is truncated to 80 values which give us acceptable accuracy with reduced complexity.}
	\begin{table}[h!]
		\centering
		\caption {Various parameters related to the FSO and RF links.}
		\label{para}
		\begin{tabular}{|ll|ll|}
			\hline
			\multicolumn{2}{|l|}{FSO link} &  \multicolumn{2}{l|}{RF link}   \\
			\hline
			\multicolumn{1}{|l|}{$\lambda_f$} & 1550 nm & \multicolumn{1}{l|}{$f_{RF}$} & 2 GHz \\
			\hline		
			\multicolumn{1}{|l|}{$D_S$} & 0.15 m & \multicolumn{1}{l|}{$\alpha_t$} & 2 \\
			\hline		
			\multicolumn{1}{|l|}{$D_r$} & 0.25 m & \multicolumn{1}{l|}{H} & 17 km \\
			\hline
			\multicolumn{1}{|l|}{$A_{atm}$} & 0.5 dB  & \multicolumn{1}{l|}{$R_n$} & 500 m \\
			\hline		
			\multicolumn{1}{|l|}{$A_{FS}$} & 268 dB & \multicolumn{1}{l|}{$B_r$} & 20 MHz \\
			\hline		
			\multicolumn{1}{|l|}{$L_{lenses}$} & 3 dB & \multicolumn{1}{l|}{$T$} & 300 k  \\
			\hline
			\multicolumn{1}{|l|}{$M_s$} & 3 dB & \multicolumn{1}{l|}{} & \\
			\hline		
			\multicolumn{1}{|l|}{$B_o$} & 30 GHz & \multicolumn{1}{l|}{} &   \\
			\hline
		\end{tabular}
		\vspace{-1em}
	\end{table}
	To show the impact of number of transmit antennas and the fading severity,  different values of $N_t$ and $m$ are considered for the RF link. Further, various values of $N$ and $\rho$ are considered to highlight the impact of user selection and the channel outdatedness on the system performance.  Furthermore, $\delta_{th}=\gamma_{th}= 5$ dB is considered for the analysis.
	\begin{figure*}
		\hfill
		\subfigure[{Heterodyne detection with negligible pointing error}]{\includegraphics[width=3.1in,height=2.6in]{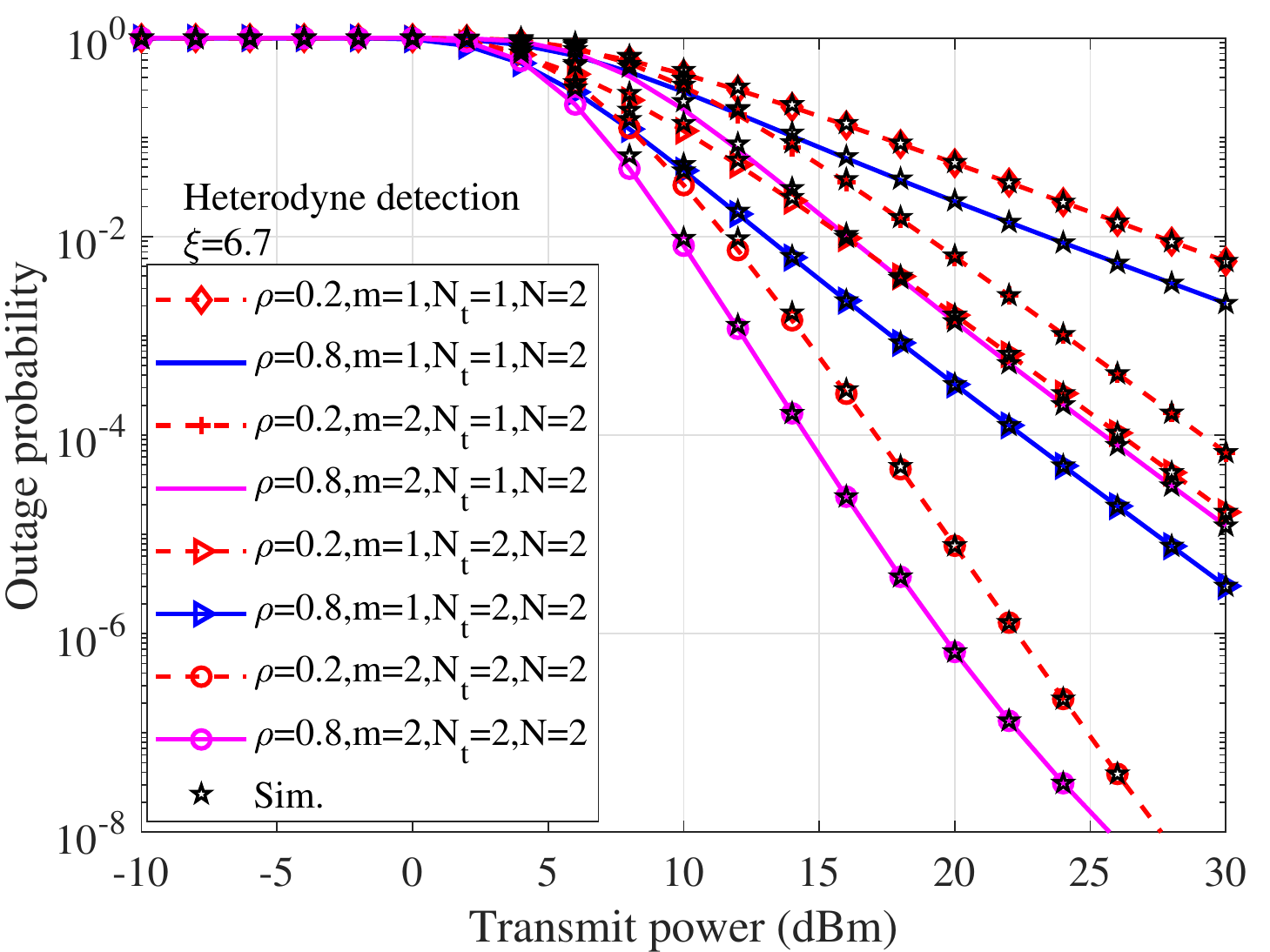}
			\label{a}}	
		\hfill
		\subfigure[{Heterodyne detection with severe pointing error}]{\includegraphics[width=3.1in,height=2.6in]{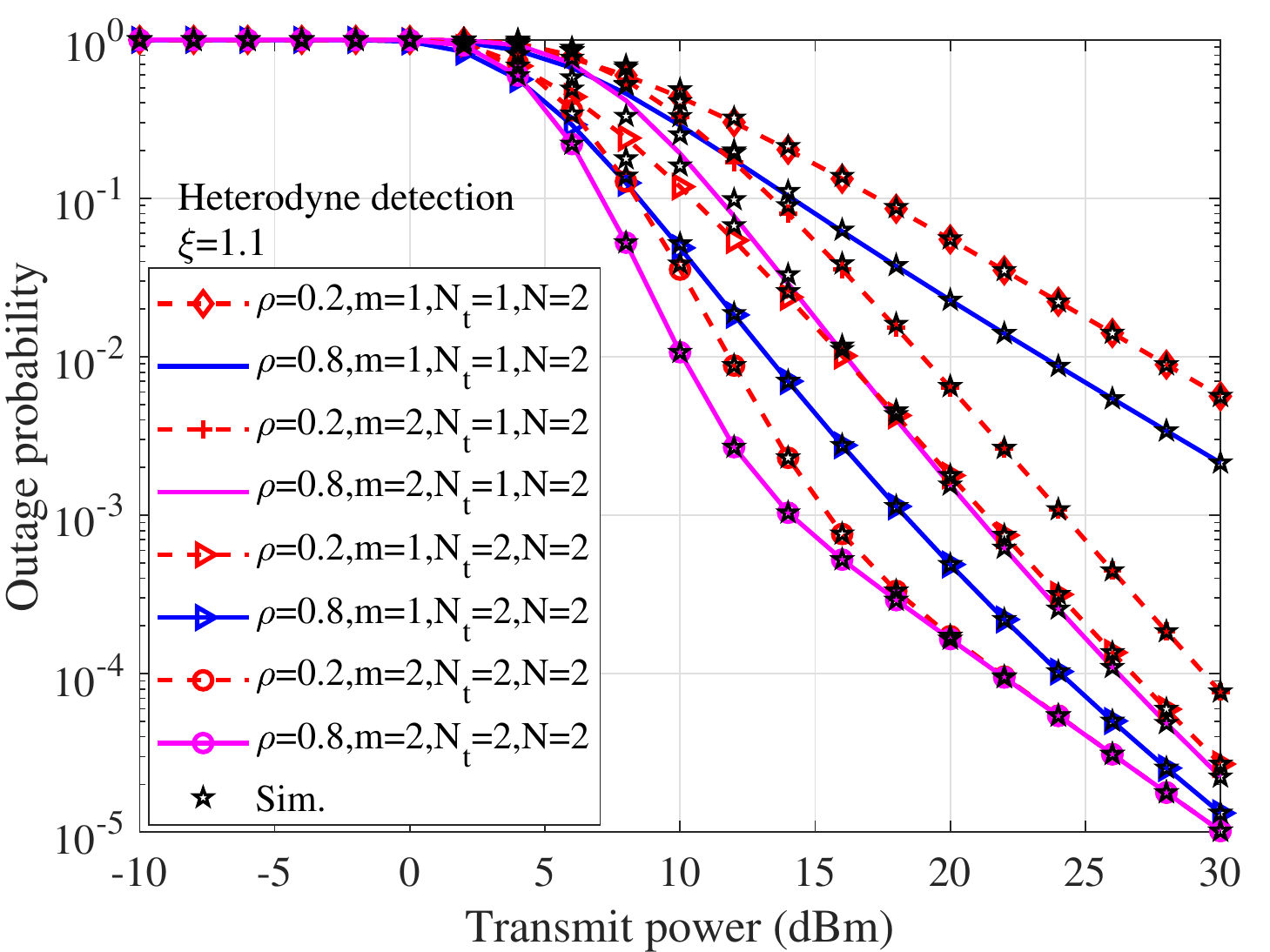}
			\label{b}}
		\hfill
		\caption{Comparison of theoretical and simulation results of outage probability versus transmit power.}
		\label{Outage}	
		\hrule
	\end{figure*} 
	
	In Fig. \ref{Outage}, theoretical and simulation results of outage probability are compared while heterodyne detection with negligible pointing error ($\xi=6.7$) and  severe pointing error ($\xi=1.1$) are considered, respectively. Further, $N=2$ is fixed and the impact of $N_t$, $m$, and $\rho$ are observed on the performance. In Fig. \ref{a} with $\xi=6.7$, a clear improvement in the outage performance is observed by increasing $\rho$ from 0.2 to 0.8  for all the investigated cases as $\rho=1$ corresponds to the perfect CSI case. To achieve an outage probability of $10^{-4}$, with the increase in $\rho$, nearly 3.7 dB gain is obtained for all the investigated cases except the case when $m=2, N_t=2, N=2$, where a gain of $2.6$ dB is obtained. The gain is reduced due to the fact that with $m=2, N_t=2$ the system reaches to its maximum performance because the performance is limited to the minimum of both the link's SNR. Further, it is observed that the increase in $N_t$ provides more gain as compared to the increase in $m$. Considering $\rho=0.8, m=1, N_t=1$ as a reference case, approximately 3 dB gain is observed with the increase in $N_t$ from 1 to 2 as compared to the increase in $m$ from 1 to 2 to obtain $10^{-4}$ outage probability. Monte-Carlo simulations are performed for all the investigated cases which validate the correctness of all the theoretical results.
	In Fig. \ref{b} with $\xi=1.1$, a clear improvement in the outage performance is observed with the increase in $\rho$ from 0.2 to 0.8  for all the cases except for $m=2, N_t=2, N=2$ throughout the transmit power range and similar trends can be seen as seen in Fig. \ref{a}. For $m=2, N_t=2, N=2$, with the increase in $\rho$, the improvement in outage performance is observed only till mid transmit power range. At high transmit powers, the performance is limited as it is reached to the maximum performance and no significant performance improvement can be observed. This is due to the fact that the FSO link's performance is limited due to the severe pointing error which limits the overall system performance. Further, the simulations validate the derived theoretical results for all the cases.
	\begin{figure}
		\centering
		\includegraphics[width=5in,height=3.3in]{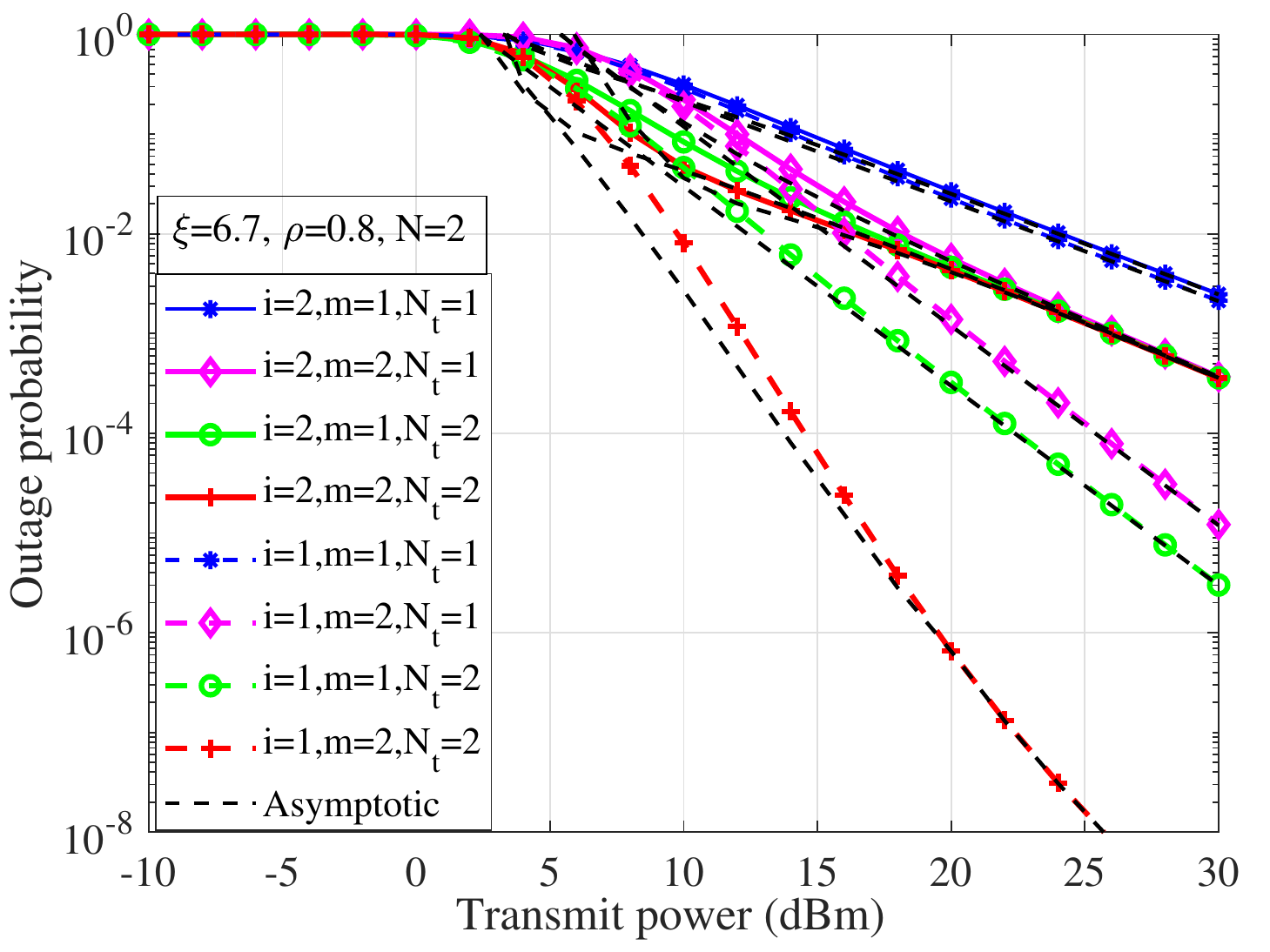}
		\caption{{Comparison of theoretical results of outage and asymptotic outage probability for both the IM/DD and heterodyne detection techniques.}}
		\label{c}
	\end{figure}
	
	In Fig. \ref{c}, theoretical results of outage probability are compared for both the IM/DD ($i=2$) and heterodyne detection ($i=1$) techniques. For the analysis, $\xi=6.7,\rho=0.8, N=2$ are fixed and the impact of $m$ and $N_t$ are observed for both the detection techniques. It is observed, that the heterodyne detection  outperforms the IM/DD technique for all the investigated cases because the heterodyne detection handles the turbulence effects in more efficient manner despite its implementation complexity \cite{zedini2016performance}. Considering the IM/DD technique and $m=1, N_t=1$ as a reference case, considerable improvement in outage performance is observed only till the mid transmit power range with the increase in $m$ or $N_t$ or both. No considerable improvement at high transmit powers is noticed  because the performance is dominated with the FSO link. On the other hand, for heterodyne technique with $m=1, N_t=1$ as a reference case, for an outage of $10^{-2}$ order, significant performance gain is observed with the increase in $m$ (7.5 dB) or $N_t$ (10.5 dB) or both (13.5 dB) which is clear from  Fig. \ref{c}. Further, to find out the diversity order of the communication system, asymptotic outage probability results are also obtained by performing the high SNR approximation. From Fig. \ref{c}, it can be seen that the asymptotic results match the outage probability results at high transmit powers which justifies the high SNR analysis. The diversity order of the considered system is $\min(\frac{\xi^2}{i},\frac{\alpha}{i},\frac{\beta}{i}, NmN_t)$ for the perfect CSI case which depends upon the atmospheric turbulence, pointing error, FSO detection type, number of users, fading parameter, and number of transmit antennas. However, with outdated CSI, the impact of user selection is vanished and the diversity order reduced to $\min(\frac{\xi^2}{i},\frac{\alpha}{i},\frac{\beta}{i}, mN_t)$.
	\begin{figure*}
		\centering
		\includegraphics[width=7in,height=2.4in]{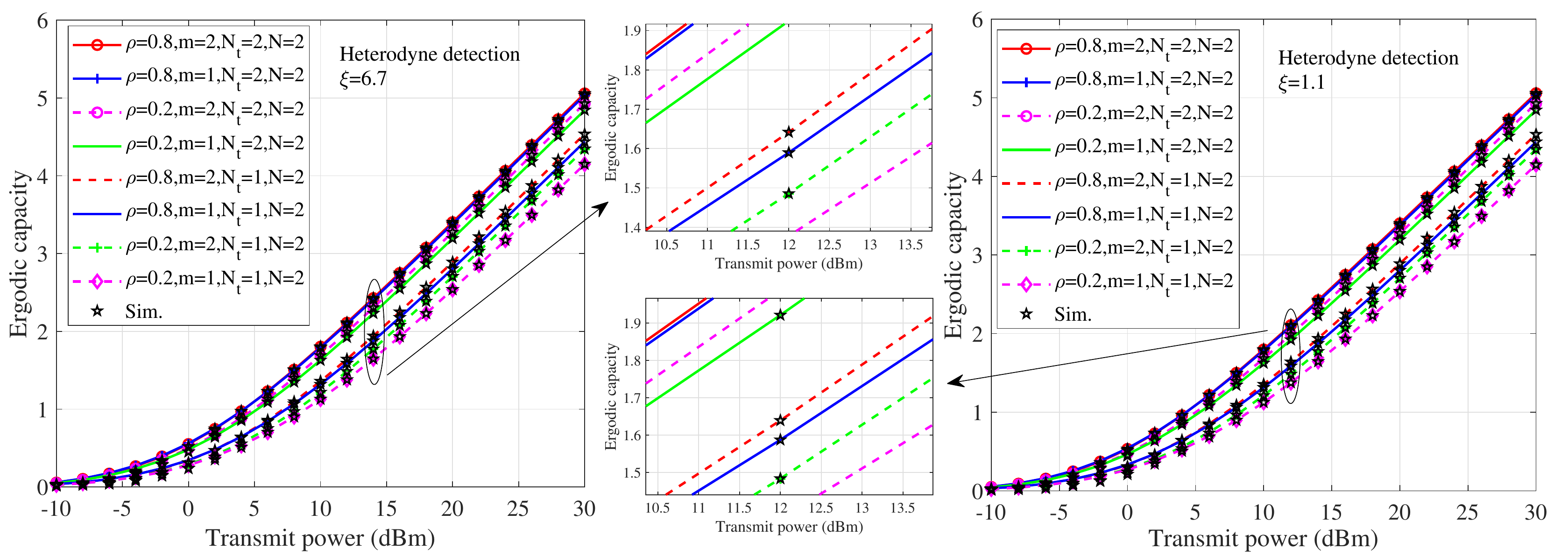}
		\caption{Comparison of theoretical and simulation results of ergodic capacity versus transmit power.}
		\label{cap}
		\hrulefill
	\end{figure*}
	
	In Fig. \ref{cap}, theoretical and simulation results of ergodic capacity  are compared while heterodyne detection with negligible pointing error ($\xi=6.7$) and  severe pointing error ($\xi=1.1$) are considered, respectively. Further, $N=2$ is fixed and the impact of $N_t$, $m$, and $\rho$ are observed on the performance. 
	In Fig. \ref{cap} (left), a clear improvement in the ergodic capacity is observed with the increase in $\rho$ from 0.2 to 0.8  for all the investigated cases. Further, it is observed that the increase in $N_t$ improves the performance more as compared to the increase in $m$. Let us consider the case  $\rho=0.2, m=1, N_t=1, N=2$ as benchmark, and fixing the transmit power to 20 dBm, it is observed that with the increase in $m$ from 1 to 2, a gain of 0.18 bits/s/Hz is obtained in the ergodic capacity. On the other hand, with the increase in $N_t$ from 1 to 2, a gain of 0.67 bits/s/Hz is obtained. Further, with the increase in $\rho$ from 0.2 to 0.8, a gain of 0.26 bits/s/Hz is obtained.
	Similar observations can be seen for the severe pointing error case as shown in Fig. \ref{cap} (right), however, a little decrement in performance is observed for all the investigated cases. Magnified portions of both the figures are also highlighted to show the change in performance with the above considered parameters ($\rho, m, N_t$).   
	Further, all the theoretical results present in Fig. \ref{cap}  are validated through the simulation results.
	
	\begin{figure}[]
		\centering
		\includegraphics[width=5in,height=3.3in]{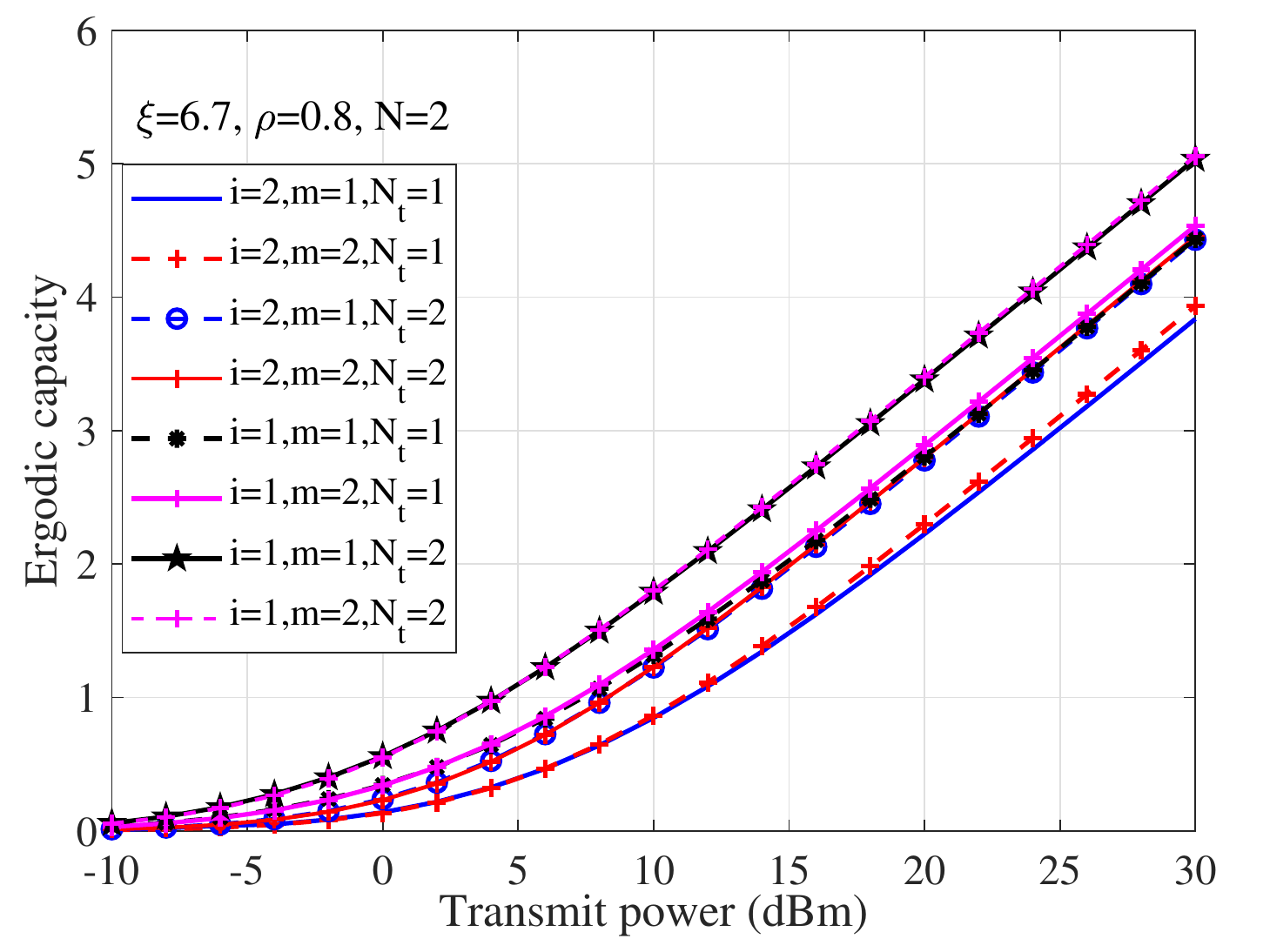}
		\caption{Comparison of theoretical results of ergodic capacity for both the IM/DD and heterodyne detection techniques.}
		\label{capc}
	\end{figure}
	In Fig. \ref{capc}, theoretical results of ergodic capacity are compared for both the IM/DD and heterodyne detection techniques. For the analysis, $\xi=6.7,\rho=0.8, N=2$ are fixed and the impact of $m$ and $N_t$ are observed for both the detection techniques.  From Fig. \ref{capc},  a significant gain in ergodic capacity is observed for the heterodyne detection as compared to the IM/DD technique throughout the transmit power range. Fixing transmit power at 15 dBm, nearly 0.55 bits/s/Hz gain in capacity is observed for the heterodyne detection as compared to the IM/DD  for all the considered cases.
	Considering $m=1, N_t=1$ as the reference case, considerable improvement in ergodic capacity is obtained with the increase in $m$ or $N_t$ or both for both the detection techniques. However, with the increase in $N_t$, ergodic capacity improve more as compared to the increase in $m$. Further, negligible improvement in the ergodic capacity is observed while moving from  $m=1, N_t=2$ to $m=2, N_t=2$ for both the detection techniques.
	This is due to the fact that the overall system capacity reached to its maximum value as the performance is limited by the weaker link (FSO link in this case).
	
	\begin{figure}[]
		\centering
		\includegraphics[width=5in,height=3.3in]{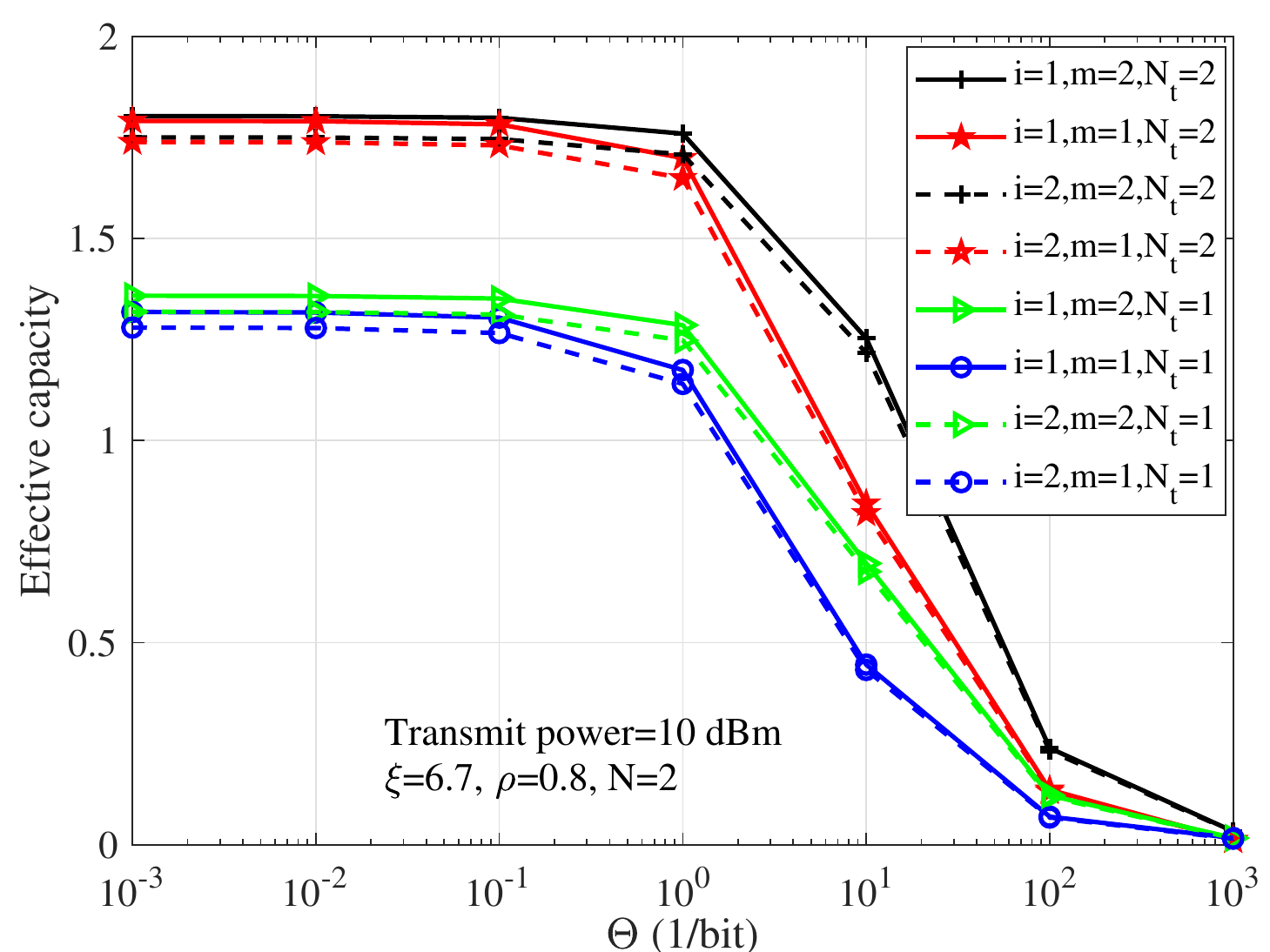}
		\caption{Effective capacity versus delay constraint $\Theta$ for both the IM/DD and heterodyne detection techniques.}
		\label{EffCap}
	\end{figure}
	In Fig. \ref{EffCap}, a comparison of effective capacity against the delay constraint $\Theta$ is shown for both the IM/DD and heterodyne techniques. For the analysis, $\xi=6.7$, $\rho=0.8$, and $N=2$ are fixed and the impact of $m$, $N_t$, and detection techniques are highlighted. Further, transmit power is fixed to 10 dBm for analysis. From Fig. \ref{EffCap}, it is observed that effective capacity improves with the increase in $m$ or $N_t$ or both. Also, effective capacity is better for $m=1, N_t=2$ case as compared to the $m=2, N_t=1$ case for both the detection techniques.  For all the combinations of $m$ and $N_t$, effective capacity for the heterodyne detection is better than the IM/DD. Further, it is observed that the effective capacity reduces with the increase in delay constraint $\Theta$ and becomes almost 0 for very high values of $\Theta$ for both the detection techniques.
	\begin{figure}
		\centering
		\includegraphics[width=5in,height=3.3in]{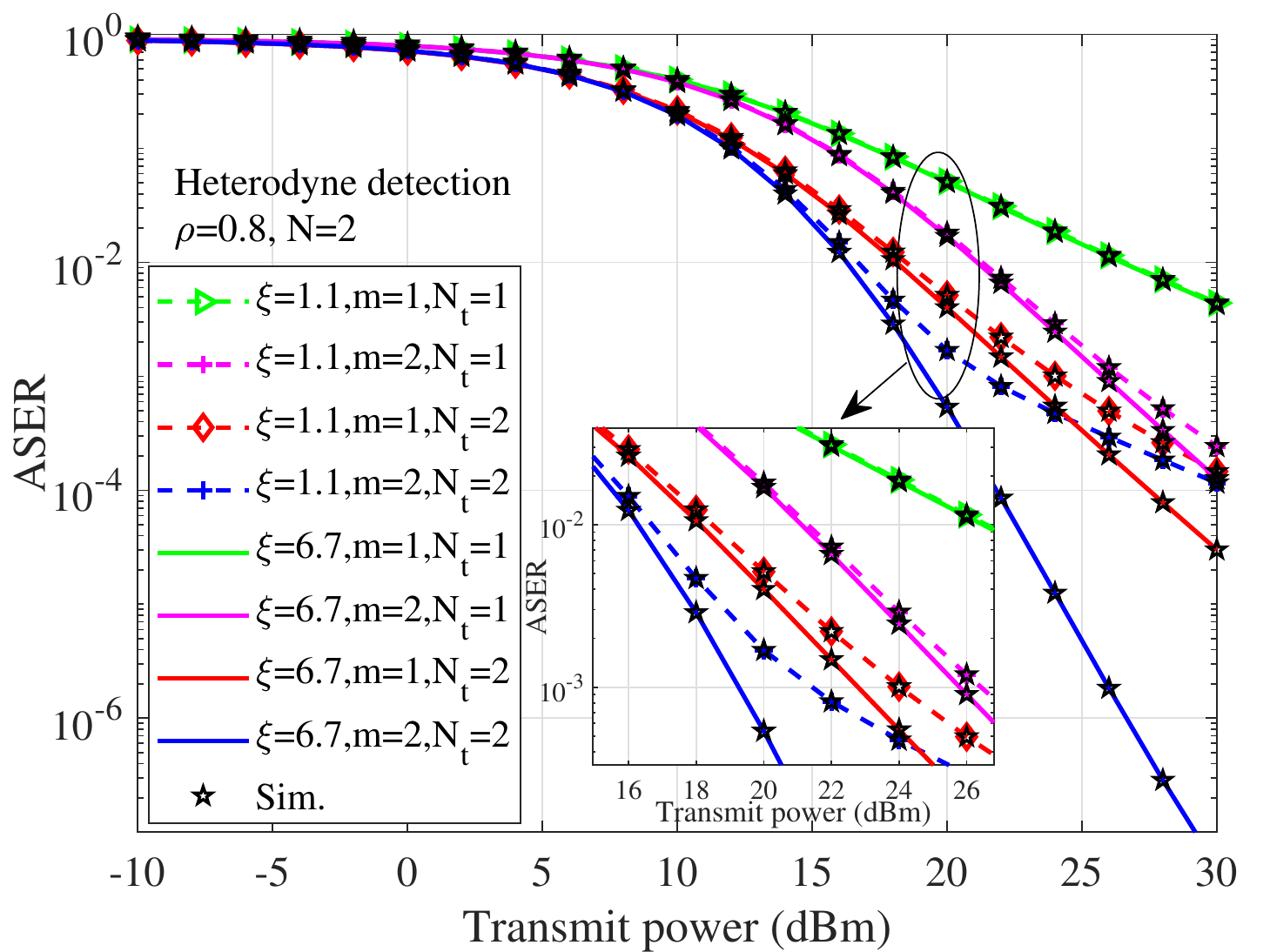}
		\caption{Theoretical and simulation results of 16-HQAM  for heterodyne detection.}
		\label{HQAM16}
		\vspace{-1em}
	\end{figure}
	\begin{figure}
		\centering
		\includegraphics[width=5in,height=3.3in]{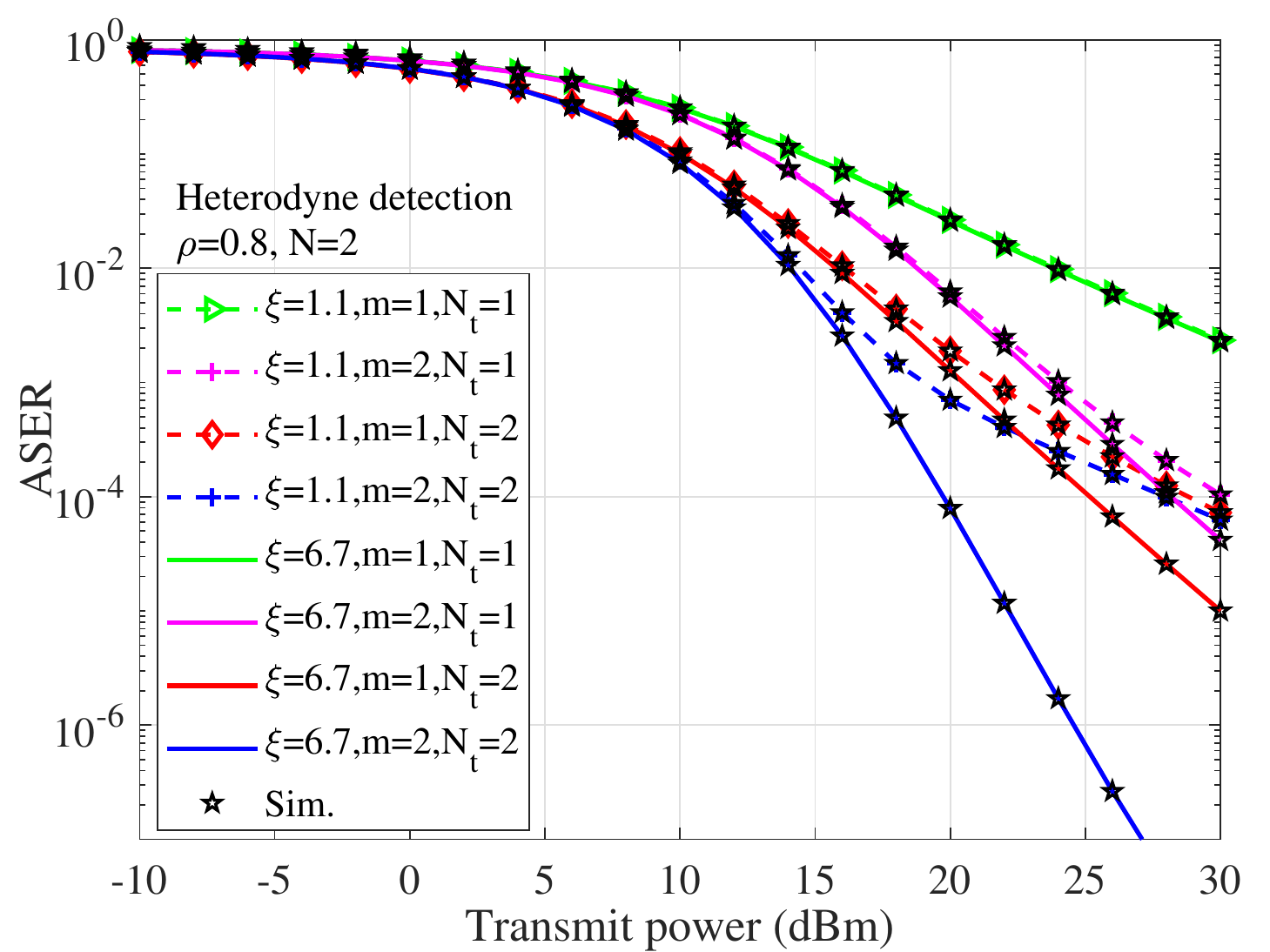}
		\caption{Theoretical and simulation results of $4 \times 2$-QAM  for heterodyne detection.}
		\label{RQAM8}
		\vspace{-1em}
	\end{figure}
	
	In Fig. \ref{HQAM16}, theoretical and simulation results of 16-HQAM are compared for various cases. For the analysis, heterodyne detection is considered at the FSO receiver and $\rho=0.8$ is fixed for the analysis.  Further, the impact of $\xi$, $m$, and $N_t$ are shown on the ASER performance for $N=2$ users.
	It can be seen that the ASER performance improves significantly by changing  $\xi=1.1$ (severe pointing error) to $\xi=6.7$ (negligible pointing error). Considering $\xi=1.1$ and $m=1, N_t=1, N=2$ as the reference case, it is observed that the performance improved significantly with the increase in $m$ or $N_t$ or both in the low and medium transmit power regimes. However, no significant improvement in ASER performance is observed at high transmit powers. This is due to the fact that there exist a severe pointing error in FSO link which limits the performance of the FSO link, and hence, the overall system performance is dominated by the FSO link. On the other hand, considering $\xi=6.7$ and $m=1, N_t=1, N=2$ as the reference case, we observe a significant improvement in ASER performance with the increase in $m$ or $N_t$ or both, throughout the transmit power range. To achieve an ASER of $10^{-2}$ order, we obtain approximately 5.4 dB, 8.4 dB, and 10.5 dB respective gain when increasing $m$ from 1 to 2, or $N_t$ from 1 to 2, or both. All the theoretical results shown in Fig. \ref{HQAM16} are validated through the simulations. The results shown in Fig. \ref{HQAM16} consider $\rho=0.8$, however, with $\rho=0.2$, similar outcomes will come with little decrement in the system performance.
	
	In Fig. \ref{RQAM8}, theoretical and simulation results of $4\times2$-QAM are compared for various cases. For the analysis, heterodyne detection is considered at the FSO receiver and $\rho=0.8$ is fixed for the analysis.  Further, the impact of $\xi$, $m$, and $N_t$ are shown on the ASER performance for $N=2$ users. It is observed that the ASER performance improves significantly with the change in  $\xi$ from 1.1 to 6.7.
	Considering $\xi=1.1$ and $m=1, N_t=1, N=2$ as the reference case, significant performance improvement is observed with the increase in $m$ or $N_t$ or both in the low and medium transmit power regimes. At  high transmit powers, no significant improvement in ASER performance is observed due to the severe pointing error in FSO link which limits the performance of the FSO link, and hence, the overall system performance is dominated by the FSO link.  On the other hand, considering $\xi=6.7$ and $m=1, N_t=1, N=2$ as the reference case,  a significant improvement in ASER performance is observed  throughout the transmit power range with the increase in $m$ or $N_t$ or both. All the theoretical results shown in Fig. \ref{RQAM8} are validated through the simulations. The results shown in Fig. \ref{RQAM8} consider $\rho=0.8$, however, with $\rho=0.2$, similar outcomes will come with little decrement in the system performance.

	\begin{figure}
		\centering
		\includegraphics[width=5in,height=3.3in]{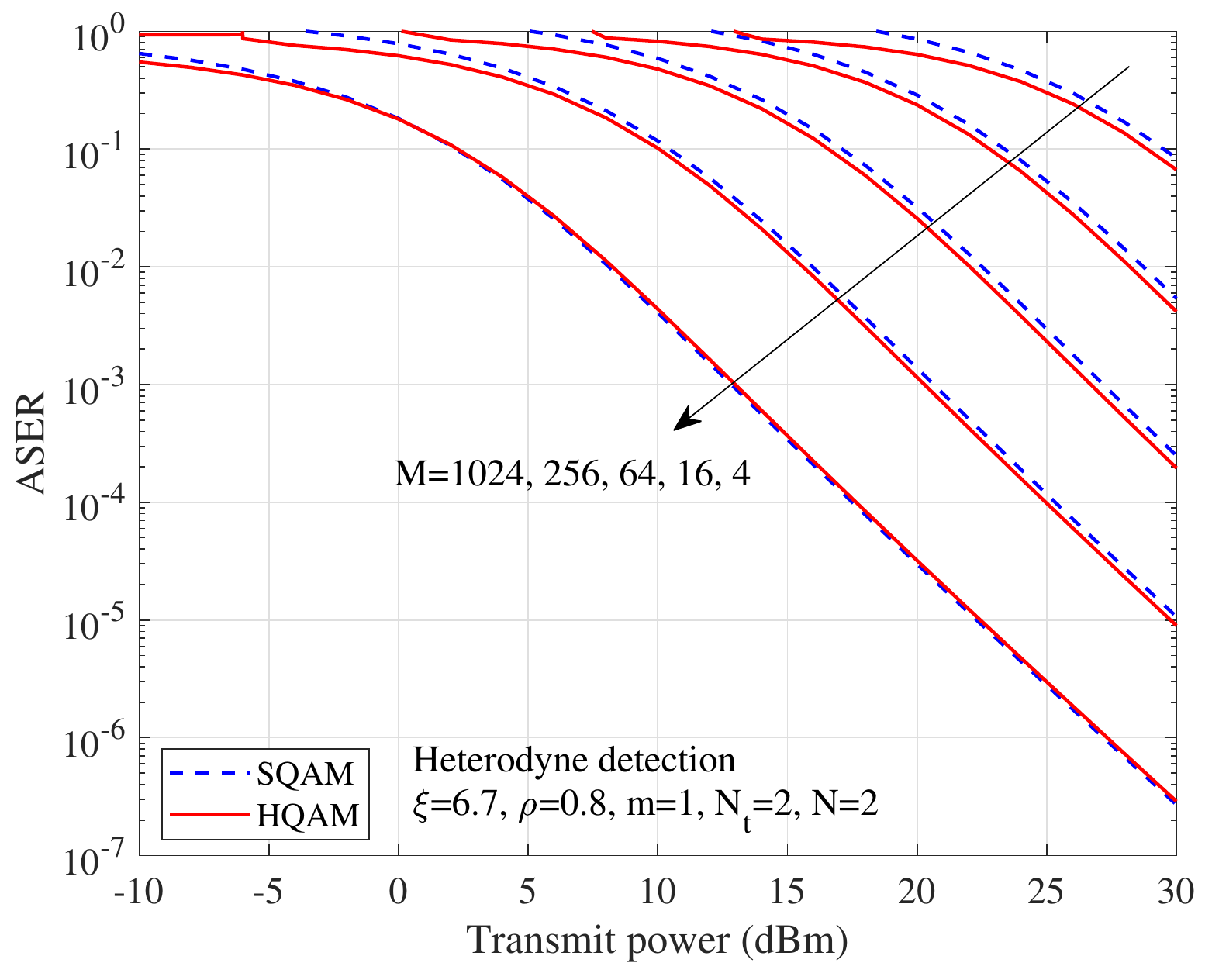}
		\caption{Comparison of the theoretical results of even power of 2 QAM constellations.}
		\label{HQAM_SQAM}
		\vspace{-1em}
	\end{figure}
	
	\begin{figure}
		\centering
		\includegraphics[width=5in,height=3.3in]{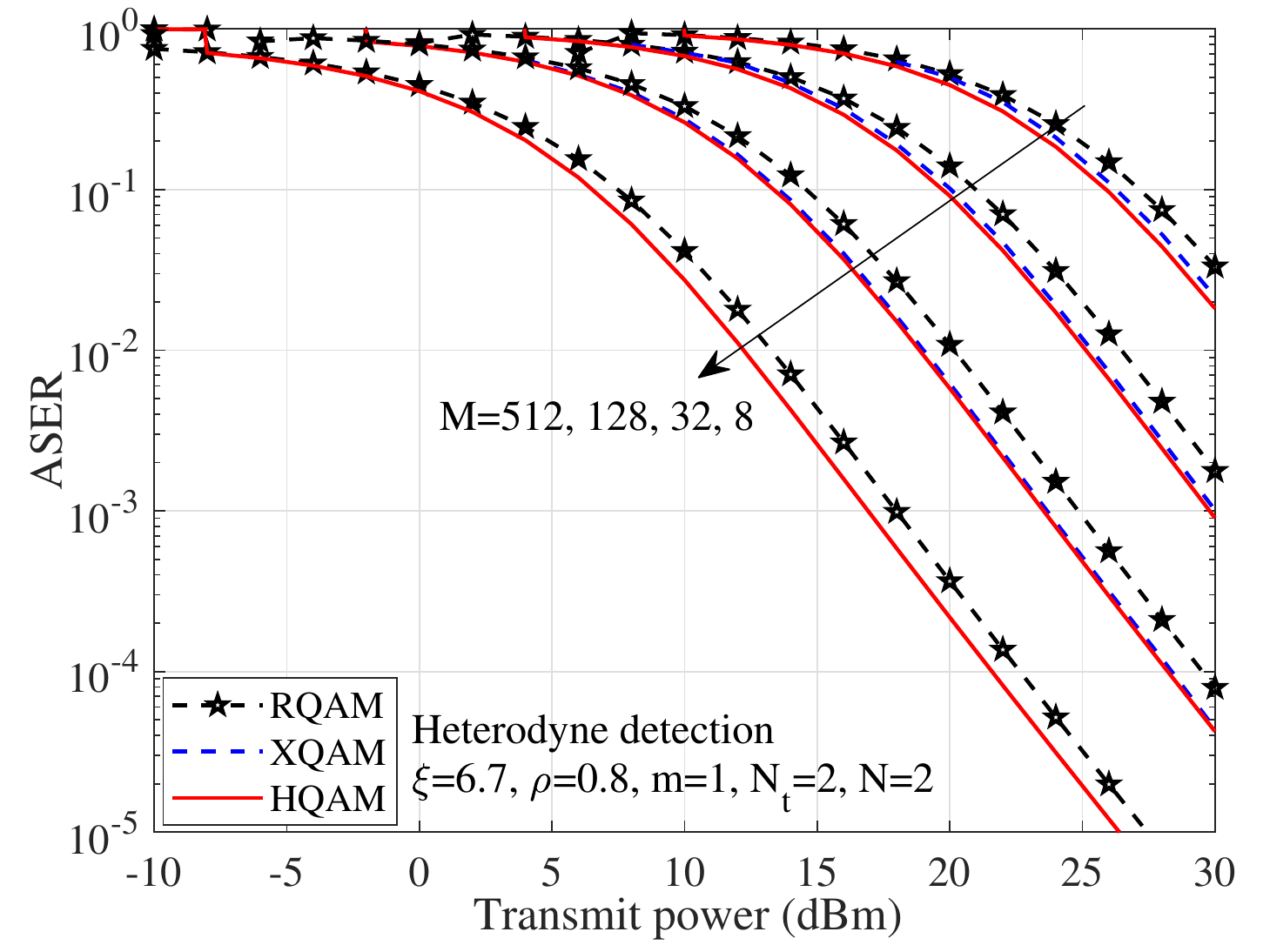}
		\caption{Comparison of the theoretical results of odd power of 2 QAM constellations.}
		\label{HXQAM}
		\vspace{-1em}
	\end{figure}
	In Fig. \ref{HQAM_SQAM}, theoretical results of even power of 2 QAM constellations are compared up to a constellation size 1024. For the the analysis $\xi=6.7$, $\rho=0.8$, $m=1$, $N_t=2$, and $N=2$ are considered. It is observed that the performance of the HQAM is better than the SQAM for all the constellation size except for $M=4$. This is because both the constellations have same average power but 4-SQAM has less number of nearest neighborhood than the 4-HQAM. For an ASER of $10^{-3}$, 4-SQAM provides nearly 0.14 dB gain over 4-HQAM. Further, with the increase in $M$, the HQAM performance improves accordingly which is clear from Fig. \ref{HQAM_SQAM}. For an ASER of $10^{-3}$, 16-HQAM, 64-HQAM, and 256-HQAM,  provide nearly 0.3 dB, 0.5 dB, 0.65 dB gain over respective SQAM which will further increase with the increase in constellation order.
	
	In Fig. \ref{HXQAM}, theoretical results of odd power of 2 QAM constellations are compared. For the the analysis $\xi=6.7$, $\rho=0.8$, $m=1$, $N_t=2$, and $N=2$  are considered. It is observed that the XQAM provides significant performance gain than the RQAM for all the constellation orders. This is because the XQAM has low peak and average powers as compared to the RQAM due to its more compact design than the RQAM. Further, HQAM provides considerable SNR gain over the XQAM which increases with the increase in M. Hence, from the above results, it is justified that the HQAM is the optimum constellation than others.

	\section{Conclusions}
	In this work, a multiuser multi-antenna UAV assisted DF based downlink terrestrial-satellite communication system was considered which operates over the mixed FSO/RF channels. Gamma-Gamma distribution with pointing error impairments was considered for the FSO channel modeling and both the IM/DD and heterodyne detection techniques were studied. Nakagami-m distribution was preferred for the modeling of RF links. Opportunistic user scheduling was performed for the terrestrial users to attain the multiuser diversity and channel outdatedness was considered during the user selection and signal transmission phase. For the performance analysis, outage probability, asymptotic outage probability, ergodic capacity, effective capacity, and ASER results of HQAM, RQAM, and XQAM constellations were obtained. The impact of pointing error impairment, FSO detector, channel outdatedness, number of users, number of transmit antennas, fading severity of RF links, and delay constraint were highlighted on all the performance metrics. Finally, all the results were validated with Monte-Carlo simulations. From the diversity analysis  it was concluded that in case of outdated CSI, no impact of user selection is observed on the system performance. Increase in number of antennas provides improved performance than the increase in fading severity. It was also observed that the  heterodyne detection  outperforms the IM/DD technique because it handles the turbulence effects more efficiently despite its implementation complexity. Further, it was observed that the effective capacity reduces with the increase in delay constraint and becomes almost zero for the high values of delay constraint. From the ASER results it was concluded that the HQAM provides improved performance than the other QAM schemes and the performance improves further with the increased constellation orders. This justifies HQAM as the optimum constellation.
	
	\appendix
	{Meijer-G is the key function in the expression of outage probability because the satellite to UAV link (FSO link) is Gamma-Gamma distributed. 
		At the high SNR, the Meijer-G function can be approximated in (\ref{OutF})  in terms of the basic elementary
		function as  \cite[(07.34.06.0001.01)]{wolframe}
		\begin{align} 
			\text{G}^{a,b}_{c,d}\Big[x\Big{|}^{\tau_3}_{\tau_4}\Big]& \approx \sum_{p=1}^{a}x^{\tau_4,p}\frac{\prod_{\underset{q\neq p}{q=1}}^{a} \Gamma(\tau_{4,q}-\tau_{4,p})\prod_{\underset{q=1}{}}^{b}\Gamma(1-\tau_{3,q}+\tau_{4,p})}{\prod_{\underset{q=b+1}{}}^{c}\Gamma(\tau_{3,q}-\tau_{4,p})\prod_{\underset{a+1}{}}^{d}\Gamma(1-\tau_{4,q}+\tau_{4,p})},
			\label{FSOapp}
		\end{align} 
		where $\tau_3=[1,\tau_1]$ and $\tau_4=[\tau_2,0]$. From (\ref{FSOapp}), it is observed that the dominant term of meijer-G function is $\min(\frac{\xi^2}{i},\frac{\alpha}{i},\frac{\beta}{i})$.}
	
	{To obtain the approximate CDF expression of the terrestrial link for the perfect CSI case ($\rho=1$), first the PDF and CDF of the SNR of the terrestrial link (\ref{CPDF_Nak}) are approximated at high SNR by considering the high SNR approximation of $\Upsilon(m,x)\underset{\overset{x=0}{}}{\approx}\frac{x^m}{m}$. Thus, the approximate PDF and CDF are respectively given as $f_{\gamma_{RU_n}}(x)=\frac{1}{\Gamma(mN_t)}\Big(\frac{m}{\bar{\gamma}_{U}}\Big)^{mN_t}x^{mN_t-1}$ and $F_{\gamma_{RU_n}}(x)=\frac{1}{\Gamma(mN_t)}\Big(\frac{m}{\bar{\gamma}_{U}}x\Big)^{mN_t}$. Substituting them in (\ref{order}), the approximate CDF expression of the terrestrial link for the perfect CSI case is obtained as (\ref{RF_per}).} 
	
	{To obtain the approximate CDF expression of the terrestrial link for the outdated CSI case ($\rho<1$), first (\ref{cond}) is approximated by considering the high SNR approximation of $I_v(x)=\frac{1}{\Gamma(v+1)}(\frac{x}{2})^v$, and then is substituted in (\ref{outd}) along with $f_{\gamma_{RU}}(x)$ (\ref{GRU}). Thus,  $f_{\hat{\gamma}_{RU}}(x)$ is calculated as
		\begin{align}\label{outd1}
			f_{\hat{\gamma}_{RU}}(x)&\approx N\sum_{k=0}^{N-1}\binom{N-1}{k}\frac{(-1)^k}{(\Gamma(mN_t))^2}\sum_{l=0}^{k(mN_t-1)}\varphi_l^k
			\Big(\frac{m}{\bar{\gamma}_{U}(1-\rho)}\Big)^{mN_t}\Big((k+1)+\frac{\rho}{1-\rho}\Big)^{-(mN_t+l)}
			\nonumber\\&
			\times
			\Gamma(mN_t+l)~	x^{mN_t-1}e^{-\frac{m}{\bar{\gamma}_{U}(1-\rho)}x}.	
		\end{align}
		Performing  $F_{\hat{\gamma}_{RU}}(\gamma_{th})=\int_{0}^{\gamma_{th}}	f_{\hat{\gamma}_{RU}}(x) dx$, and using the high SNR approximation of $\Upsilon(m,x)\underset{\overset{x=0}{}}{\approx}\frac{x^m}{m}$, the approximate CDF expression of the terrestrial link for the outdated CSI case is obtained as (\ref{RF_out}). 
	}
	
	\bibliographystyle{IEEEtran}
	\tiny
	\bibliography{ref}

	\begin{IEEEbiography}[{\includegraphics[width=1in,height=1.5in,clip,keepaspectratio]{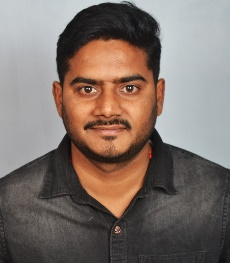}}]{Praveen K. Singya} (M'20)  received his B.E. in Electronics and Communication Engineering from Jabalpur Engineering College, Jabalpur, India in 2012. He received his M.Tech. in Communication System Engineering from VNIT, Nagpur, India in 2014. He received his Ph.D. degree from the Indian Institute of Technology Indore, India, in 2019. He is currently a postdoctoral fellow at King Abdullah University of Science and Technology (KAUST), Thuwal, Makkah Province, Saudi Arabia. His research interest includes design and performance analysis of various wireless networks over different fading channels, satellite communication, HAP communication, and free-space optics.
\end{IEEEbiography}

\begin{IEEEbiography}[{\includegraphics[width=1in,height=1.5in,clip,keepaspectratio]{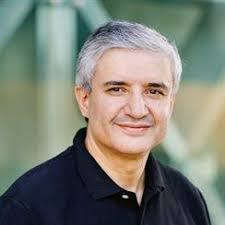}}]{Mohamed-Slim Alouini} (S’94-M’98-SM’03-F’09)  was born in Tunis, Tunisia. He received the
Ph.D. degree in Electrical Engineering
from the California Institute of Technology (Caltech), Pasadena,
CA, USA, in 1998. He served as a faculty member at the University of Minnesota,
Minneapolis, MN, USA, then in the Texas A\&M University at Qatar,
Education City, Doha, Qatar before joining King Abdullah University of
Science and Technology (KAUST), Thuwal, Makkah Province, Saudi
Arabia as a Professor of Electrical Engineering in 2009. His current
research interests include modeling, design, and
performance analysis of wireless communication systems.
\end{IEEEbiography}

\end{document}